\documentclass[a4paper]{article}
\usepackage[english]{babel}
\usepackage[utf8x]{inputenc}
\usepackage[T1]{fontenc}
\usepackage{subcaption}
\usepackage{authblk}
\usepackage{enumitem}
\usepackage{amsfonts}

\usepackage[a4paper,top=3cm,bottom=2cm,left=3cm,right=3cm,marginparwidth=1.75cm]{geometry}

\usepackage{amsmath}
\usepackage{graphicx}
\usepackage[colorinlistoftodos]{todonotes}

\newlength{\twosubht}
\newsavebox{\twosubbox}

\title{Delayed Dynamical Systems: Networks, Chimeras and Reservoir Computing}
\author[1,2]{Joseph D. Hart}
\affil[1]{Institute for Research in Electronics and Applied Physics, University of Maryland, College Park, MD 20742}
\affil[2]{Department of Physics, University of Maryland, College Park, MD 20742}
\author[3]{Laurent Larger}
\affil[3]{FEMTO-ST Institute/Optics Department, CNRS \& University Bourgogne Franche-Comt\'e, 15B avenue des Montboucons, 25030 Besan\c con Cedex, France}
\author[1,4]{Thomas E. Murphy}
\affil[4]{Department of Electrical and Computer Engineering, University of Maryland, College park, MD 20742}
\author[1,2,5]{Rajarshi Roy}
\affil[5]{Institute for Physical Science and Technology, University of Maryland, College Park, MD 20742}

\begin{document}
\maketitle
\begin{abstract}
We present a systematic approach to reveal the correspondence between time delay dynamics and networks of coupled oscillators.  After early demonstrations of the usefulness of spatio-temporal representations of time-delay system dynamics, extensive research on optoelectronic feedback loops has revealed their immense potential for realizing complex system dynamics such as chimeras in rings of coupled oscillators and applications to reservoir computing.  Delayed dynamical systems have been enriched in recent years through the application of digital signal processing techniques.  Very recently, we have showed that one can significantly extend the capabilities and implement networks with arbitrary topologies through the use of field programmable gate arrays (FPGAs).  This architecture allows the design of appropriate filters and multiple time delays which greatly extend the possibilities for exploring synchronization patterns in arbitrary topological networks.  This has enabled us to explore complex dynamics on networks with nodes that can be perfectly identical, introduce parameter heterogeneities and multiple time delays, as well as change network topologies to control the formation and evolution of patterns of synchrony.
\end{abstract}
\maketitle
\newpage

\section{Introduction}

Networks of coupled oscillators are dynamical systems of great interest for both basic and applied research. Networks are high-dimensional systems that can display a great variety of dynamical behaviors. Applications abound, from neuroscience \cite{bassett2017network} and gene regulation \cite{davidson2005gene} to the power grid \cite{motter2013spontaneous} and machine learning \cite{jaeger2002tutorial}. Networks have long been a fertile ground for theoretical research \cite{boccaletti2006complex}, however, experiments on large networks have proven difficult because of the necessity to create, connect, and measure a large number of independent oscillators. In the few cases where experiments with large networks have been possible, it is often difficult or impossible to reconfigure the network, with a few notable exceptions \cite{hagerstrom2012experimental,tinsley2012chimera,pecora2014cluster}.

Nonlinear systems with time delayed feedback are a different type of high-dimensional dynamical system that are much easier to study experimentally. Time delays often arise when the intrinsic dynamics of a system are fast enough that the finite propagation velocity of signals must be taken into account. For example, in a semiconductor laser with time delayed feedback through an external mirror, the photon lifetime is significantly shorter than the feedback time, which can cause the laser intensity to oscillate chaotically \cite{lang1980external}. From an experimental point of view, delay systems are particularly attractive because the dimensionality of the dynamics often increases linearly with the delay \cite{farmer1982chaotic,le1987conjecture}, which is typically easy to control.



The simplest delay systems can be modeled by \cite{erneux2009applied}
\begin{equation}
\label{eq:ikeda}
\tau_L \dot{x}(t) = -x(t) + F\big(x(t-\tau_D)\big),
\end{equation}
where $\tau_L$ is the intrinsic time scale of the system, $F(x)$ is a nonlinear function of $x$ and $\tau_D$ is the time delay. Equation \ref{eq:ikeda} has been used to model systems from many different areas of science \cite{erneux2017introduction}, including physiology \cite{mackey1977oscillation}, population dynamics \cite{kuang1993delay}, and laser physics \cite{ikeda1979multiple}. Systems described by Eq. \ref{eq:ikeda} have been shown to display a wide variety of interesting behaviors, including square waves \cite{chow1992sine,weicker2012strongly}, new types of chaos (in the case that $\tau_D$ varies in time) \cite{muller2018laminar}, and spatiotemporal phenomena \cite{yanchuk2017spatio}.

Indeed, research over the last 25 years has shown that a wide variety of spatio-temporal phenomena can be observed in temporal systems with a long delayed feedback. The interpretation of dynamics in delayed systems as spatio-temporal phenomena is enabled by the space-time representation \cite{arecchi1992two}. Some of the theoretically predicted and experimentally observed spatio-temporal phenomena include defect-mediated turbulence \cite{giacomelli1994defects,yanchuk2014pattern}, coarsening \cite{giacomelli2012coarsening,javaloyes2015arrest}, domain nucleation \cite{giacomelli2013nucleation}, spatial coherence resonance\cite{marino2017pseudo}, and phase transitions \cite{faggian2018evidence}. 

Our focus in this paper is on the implementation of networks of truly identical coupled oscillators through the use of a single nonlinear delayed feedback system. This is made possible through the same space-time representation that led to the observation of other spatio-temporal phenomena in delay systems. Originally invented for the implementation of neural networks for reservoir computing in hardware \cite{appeltant2011information,larger2012photonic}, this technique for implementing networks has subsequently been adapted for basic research, such as the study of chimera states in ring networks \cite{larger2013virtual,larger2015laser} and cluster synchronization in arbitrary networks \cite{hart2017experiments,siddique2018symmetry}. This framework for implementing networks is particularly attractive because it allows for experiments on large networks without building a large number of separate physical oscillators and it allows for experiments on truly identical oscillators.  We focus on opto-electronic implementations, which are popular due to their speed, cost, and ease of implementation; however, the techniques described are applicable to other delay systems as well.

In Section 2, we introduce a basic mathematical description of a delayed feedback system through a commonly used integro-differential delay equation. Additionally, we present a less commonly used, but equivalent, description from filter theory that employs a convolution integral of the feedback signal with the impulse response that describes the bandwidth limitations of the system. This second formalism, when viewed in the space-time representation, sheds insight into how networks of oscillators can be realized with a single nonlinear system with delayed feedback. Finally, we describe one particular opto-electronic oscillator that has been a favorite of experimenters due to its reliability and ease of implementation over a wide range of parameters and time scales.

The space-time representation of delay systems is presented in Section 3. The space-time representation relies on the separation of time scales--fast dynamics and a long delay--to parameterize time as a time-like integer number that counts the number of round-trip times and a continuous, space-like variable that denotes the position within each delay. This analogy between feedback systems with a long time delay and spatio-temporal systems has allowed for a deeper understanding of many complex phenomena observed in delay systems, including defect-mediated turbulence \cite{giacomelli1994defects,yanchuk2014pattern}, coarsening \cite{giacomelli2012coarsening,javaloyes2015arrest}, domain nucleation \cite{giacomelli2013nucleation}, spatial coherence resonance\cite{marino2017pseudo}, phase transitions \cite{faggian2018evidence} and now, network dynamics.

Section 4 describes in detail how the space-time representation allows for the implementation of networks of truly identical coupled oscillators using only a single delayed feedback system. Traditional networks are spatially multiplexed: all nodes are updated simultaneously in parallel depending on their previous states. Delay feedback networks replace the spatial multiplexing of traditional networks with time multiplexing, in which the single nonlinear element serially updates the nodes, which are distributed across the delay line. The nodes are coupled together by the ``inertia,'' or finite response time, of the system, which arises from the bandwidth limitations of the components. When this filtering is time-invariant, the resulting network has cyclic symmetry. In particular, Section 4 focuses on the discrete time case; e.g., when the time delay is implemented by a digital delay line.

The use of delay networks for hardware implementations of reservoir computers is discussed in Section 5. Reservoir computing--alternatively echo state networks \cite{jaeger2002tutorial} or nonlinear transient computing \cite{martinenghi2012photonic}--is a type of neural network in which only the output connections are trained (the input and internal connections are fixed). Reservoir computers are particularly attractive because they can be trained by simple linear regression and because they are well-suited for implementation in specialized hardware. Delay networks have proven to be particularly well-suited for reservoir computing.

Section 6 extends the delay network formalism developed in Section 4 to the continuous time case (the case of analog delay lines).

Chimera states are an unexpected coexistence of spatial domains of coherence and incoherence in a system of identical oscillators with symmetric coupling \cite{kuramoto2002coexistence,abrams2004chimera}. Chimera states were particularly difficult to observe in experiments because they typically (but not always \cite{panaggio2016chimera,hart2016experimental}) occur in large networks, which are difficult to experimentally implement. Initially observed in 2012 \cite{tinsley2012chimera,hagerstrom2012experimental} a decade after their prediction, they were soon after observed in electronic \cite{larger2013virtual} and opto-electronic \cite{larger2015laser} delay systems, as presented in Section 7.

A recently developed technique \cite{hart2017experiments} that allows a network with any topology to be implemented in a delay system is described in Section 8. This technique replaces the time-invariant filters used in the original delay network implementations with a time-dependent filter. The time-dependent filter, implemented digitally with a field-programmable gate array (FPGA), extends the range of networks that can be realized from only networks with rotational symmetry to networks with completely arbitrary topology.

\section{Introduction to optoelectronic oscillators with delayed feedback}

\begin{figure}
\centering
\includegraphics[width=0.9\textwidth]{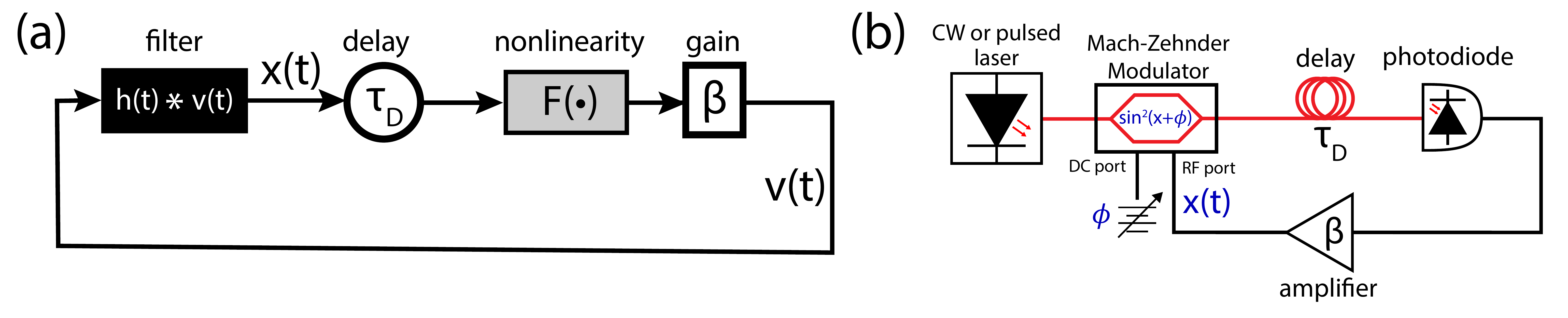}
\caption{\label{fig:MZMapparatus}Nonlinear delayed feedback system. (a) Block diagram of a delay system. $v(t)=\beta F(x(t-\tau_D))$ is the input to the linear filter described by the impulse response $h(t)$, and $x(t)$ is the filter output. (b) Experimental setup of an opto-electronic oscillator delayed feedback system. The filtering is performed either by the component with the narrowest bandwidth (usually the photodiode) or by a stand-alone filter (not shown). The oscillator can be a discrete time map (when powered by a pulsed laser) or a continuous time system (when powered by a CW laser).}
\end{figure}

The basic form of a delayed feedback system is depicted by the block diagram in Fig. \ref{fig:MZMapparatus}a. The output of a nonlinearity $F(\cdot)$ is amplified, filtered, and delayed before being fed back as the input to the nonlinearity. The filtering may either be intentionally implemented or arise from the bandwidth limitations of the system. Such a delayed feedback system can be described by the convolution of the input to the filter with the impulse response $h(t)$ that characterizes the filter \cite{oppenheim1996signals}:

\begin{equation}
\label{eq:convolution}
x(t) = h(t)*\beta F(x(t-\tau_D)) =\beta\int\limits_{-\infty}^{\infty}h(t-t')F\big(x(t'-\tau_D)\big)dt' = \beta\int\limits_{\infty}^th(t-t')F\big(x(t'-\tau_D)\big)dt'
\end{equation}
where in the last step we use the property that $h(t)$ is causal. In Eq. \ref{eq:convolution}, $x(t)$ is the filter output, $\beta$ is the round trip gain, and $\tau_D$ is the time delay. 

If the form of the filter is known, an equivalent delay differential equation can be used to describe the system. In the case that the bandwidth limitations of the system can be accurately described by a two-pole bandpass filter, the delay differential equation is

\begin{equation}
\label{eq:DDE}
\tau_L\dot{x}(t) = -\big(1+\frac{\tau_L}{\tau_H}\big)x(t) - \frac{1}{\tau_H}\int\limits_{-\infty}^t x(s)ds + \beta F\big(x(t-\tau_D)\big)
\end{equation}
where $\tau_D$ is the time delay, $\tau_L=1/2\pi f_L$ is the low pass filter response time, and $\tau_H=1/2\pi f_H$ is the high pass filter response time. Equation \ref{eq:DDE} is quite general in that it can be used to model many delayed feedback systems. Indeed, by considering the limit $\tau_H\to\infty$ (i.e., the case of a low pass instead of a band pass filter), Eq. \ref{eq:DDE} reduces to Eq. \ref{eq:ikeda}.

One experimental system of particular interest that can be accurately modeled by Eq. \ref{eq:DDE} is the opto-electronic oscillator. The opto-electronic oscillator was originally studied in bulk optics \cite{gibbs1981observation} and soon after implemented using standard telecommunications components \cite{neyer1982dynamics}. These systems have been found to be extremely rich in their dynamics, in part because they can span an enormous range of time scales \cite{peil2009routes}. They have been used to study chaotic breathers \cite{kouomou2005chaotic}, broadband chaos \cite{callan2010broadband}, network dynamics \cite{ravoori2011robustness,hart2016experimental}, and the transition from noise to chaos \cite{hagerstrom2015harvesting}. Additionally, opto-electronic oscillators are useful for a variety of applications, including the generation of high-spectral purity microwaves \cite{yao1996optoelectronic}, chaos communications \cite{goedgebuer2002optical,argyris2005chaos}, and reservoir computing \cite{larger2012photonic,martinenghi2012photonic}.

A schematic of an opto-electronic oscillator is shown in Fig. \ref{fig:MZMapparatus}b. Constant intensity light from a fiber-coupled CW laser passes through an integrated electro-optic Mach-Zehnder intensity modulator, which provides the nonlinearity $F(x)=\sin^2(x+\phi)$. The quantity $x(t)$ represents the normalized voltage applied to the intensity modulator, and $\phi$ is the normalized DC bias voltage. The time delay is implemented by an optical or electronic (not shown) delay line. The filtering is performed either by the photodiode (the component with the narrowest bandwidth) or a stand-alone analog \cite{yao1996optoelectronic} or digital \cite{murphy2010complex} filter (not shown). For a  recent review of these opto-electronic oscillators, see Ref. \cite{larger2013complexity}.

Alternatively, the system can be turned into a discrete time map by pulsing the laser at a repetition rate $f_r = N/\tau_D$ \cite{larger2005flow}. In this case, the system can be modeled as 
\begin{equation}
\label{eq:map1}
x[k] = \beta\sum_{m=-\infty}^k h[k-m]F(x[m-N])
\end{equation}
where $x[k]$ is the height of the $k^{th}$ electrical pulse applied to the modulator, $h$ is the infinite impulse response of the filter sampled at the repetition rate $f_r$. As the repetition rate $f_r\to\infty$, time becomes continuous, the sum becomes a convolution integral, and we obtain Eq. \ref{eq:convolution}. Therefore, this system allows for the study of the transition from discrete to continuous time in chaotic systems.

\section{Space-time representation}
The space-time representation of delay systems was originally motivated by the numerical treatment of delay differential equations \cite{farmer1982chaotic}. The time variable is split up into a continuous variable $\sigma$ bounded between 0 and $\tau_D$, and an independent discrete variable $n$ that counts the number of delays since the origin. Ikeda and Matsumoto \cite{ikeda1989information} were the first to consider $\sigma$ to be a ``spatial'' variable in their modeling of optical turbulence. The space-time representation was formalized and first used on experimental data by Arecchi et al. in 1992 \cite{arecchi1992two} in order to study long-time correlations on the order of one delay in a CO$_2$ laser with delayed feedback. Since then, the relationship between delay systems and spatio-temporal systems has been investigated thoroughly \cite{giacomelli1994defects,larger2013virtual,larger2015laser,faggian2018evidence}, and in many cases, equivalence has been rigorously established \cite{giacomelli1996relationship,wolfrum2006eckhaus,yanchuk2014pattern,yanchuk2015dynamical}. For a recent review, see Ref. \cite{yanchuk2017spatio}.

The space-time representation of delay systems is particularly meaningful when the delay $\tau_D$ is long compared to the time scale $t_c$ of the temporal dynamics of the system, as measured by the width of the zeroth peak in the autocorrelation \cite{yanchuk2017spatio}. In this case, there is a separation of time scales, and so it is natural to parameterize time as 

\begin{equation}
\label{eq:preSTR}
t = n\tau_D+\sigma,
\end{equation}
where $n$ is an integer that counts the number of delay times since the origin, and $\sigma$ is a continuous variable between 0 and $\tau_D$ that gives the position along the delay. As a result, $n$ is often considered to be a discrete time and $\sigma$ a continuous pseudo-spatial variable. We note that $t_c$ is a property of the dynamics and therefore depends on $\beta$ and $F(x)$ in addition to the time scales $\tau_L$ and $\tau_H$ in Eq. \ref{eq:DDE}; in practice, however, it is often the case that that $t_c\approx \tau_L$ \cite{yanchuk2017spatio}.

\begin{figure}
\centering
\includegraphics[width=\textwidth]{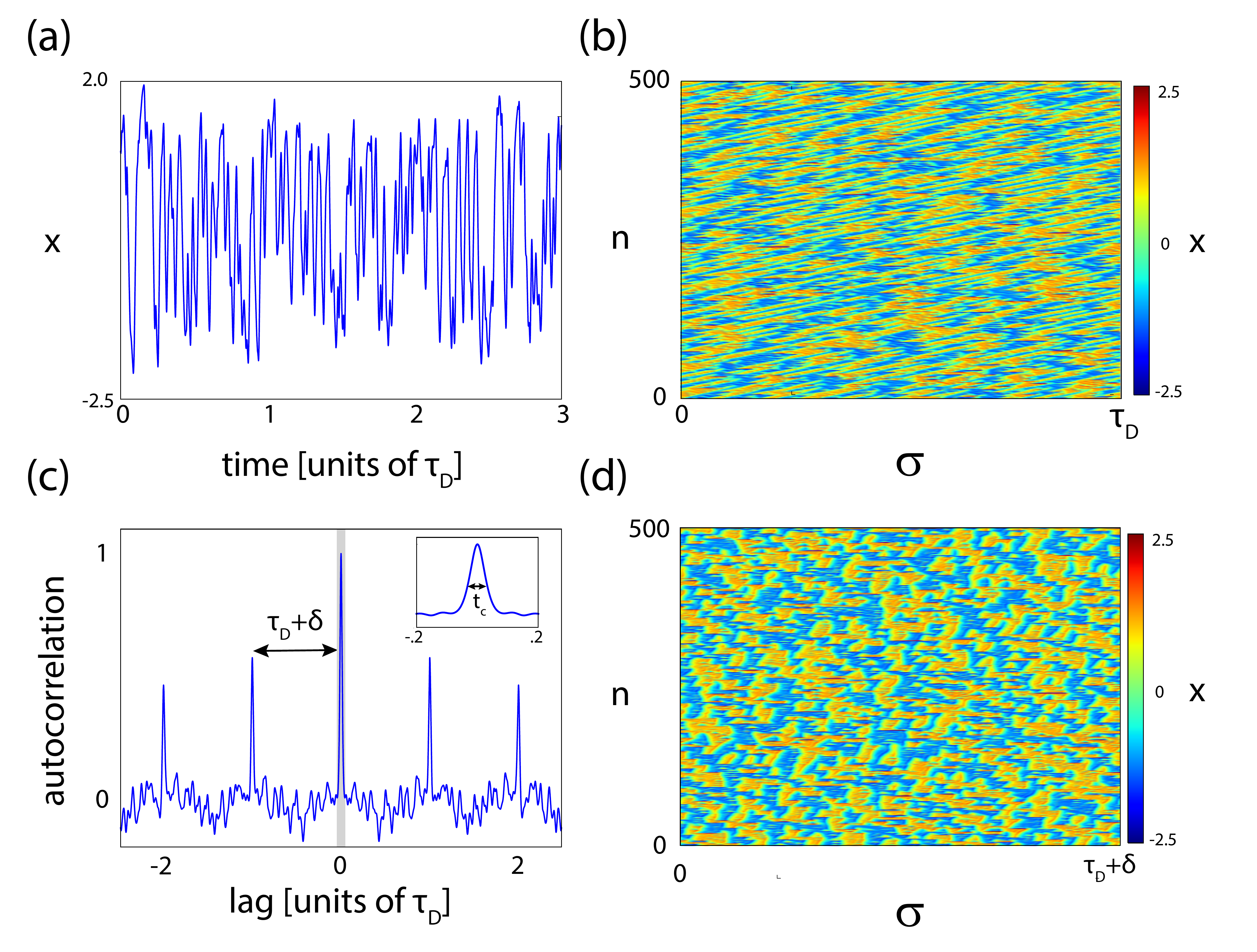}

\caption{\label{fig:STR}Illustration of the space-time representation. (a) Time series of the delayed system in Eq. \ref{eq:DDE}. (b) Space-time representation of the time series shown in (a), where $\sigma\in[0,\tau_D]$. (c) Autocorrelation of the time series shown in (a) The distance to the first autocorrelation peak is $\tau_D+\delta$. Here $\tau_D = 4$ ms and $\delta = 250 \mu$s. \textit{Inset}. Zoom in on central autocorrelation peak. The width of this peak is $t_c$. (d) Space-time representation with drift correction ($\sigma\in[0,\tau_D+\delta]$). These figures were made from a numerical simulation of Eq. \ref{eq:DDE} with $\beta = -5$, $\tau_L = 400$ $\mu$s, $\tau_H = 10$ ms, $\tau_D = 4$ ms, and $F(x)=\sin^2(x(t)-\pi/4)$, which describes the opto-electronic oscillator shown in Fig. \ref{fig:MZMapparatus}b. }
\end{figure}

When working with delay systems, one often obtains a long time series $x(t)$ such as the one shown in Fig. \ref{fig:STR}(a). It seems that there are (and indeed one expects there to be) correlations on the order of one time delay $\tau_D$. Plotting the time series in the space-time representation in Fig. \ref{fig:STR}(b) shows long time correlations (on the order of several $\tau_D$) as spatial structures that evolve in discrete time.

While Fig. \ref{fig:STR}(b) does reveal long-time correlations as spatio-temporal structures, it is clear that as $n$ increases the structures are drifting to the right in $\sigma$-space. In other words, the long-time correlations occur over a time slightly larger than $\tau_D$. This can be seen by looking at the autocorrelation of the time series, shown in Fig. \ref{fig:STR}(c). The autocorrelation begins to increase near a lag of $\tau_D$, but only reaches its peak at $\tau_D+\delta$ due to the finite response time of the system \cite{yanchuk2017spatio}. Therefore $\delta$ is related to the widths of the zeroth autocorrelation peak $t_c$ as well as the width of the first autocorrelation peak. Previous works have extensively studied this drift and its relation to co-moving Lyapunov exponents \cite{giacomelli1996relationship,giacomelli2012coarsening}.

The drift is a reflection of the fact that the system is causal. The delayed term $x(t-\tau_D)$ cannot affect the dynamics before, \textit{or even at}, the time $t$. Therefore, in Fig. \ref{fig:STR}(d), we use

\begin{equation}
\label{eq:STR}
t=nT+\sigma
\end{equation}
to create space-time representations, where $T=\tau_D+\delta$ is the recurrence time and now $\sigma\in[0,T]$. When the space-time representation is done in this way, the structures are stabilized in space (i.e., they have a nearly stationary average spatial position). Indeed, it has been shown that this is often the correct moving frame in which to view the spatio-temporal behavior of time-delayed systems \cite{yanchuk2017spatio}.

\section{Using the space-time representation to realize coupled oscillators in a single delay system}
Recently, the space-time representation has been used to interpret a single nonlinear node with delayed feedback as a network of coupled oscillators. These experiments replace the spatial multiplexing of a traditional network (in which all nodes are updated simultaneously in parallel) with time multiplexing, in which the single nonlinear element serially updates each of the nodes, which are distributed across the delay line. There are two major benefits to this network implementation: this is the \textit{only} way to create a network of truly identical nodes, and it allows one to implement a large network without building a large number of separate physical nodes. While originally used for a hardware implementation of reservoir computing \cite{appeltant2011information,larger2012photonic,martinenghi2012photonic,paquot2012optoelectronic,brunner2013parallel,antonik2017brain}, these types of delay systems have since been used to study chimera states in cyclic networks \cite{larger2013virtual,larger2015laser} and cluster synchronization in arbitrary networks \cite{hart2017experiments,siddique2018symmetry}.

Because delay systems require a continuous function to describe their initial conditions, they are considered infinite dimensional systems. However, it was noticed early on that chaotic attractors of delay systems have finite dimension in practice \cite{farmer1982chaotic}. In trying to explain this finite dimensionality, Le Berre et al. conjectured that the dimension of the attractor is equal to $\tau_D/t_c$, where $t_c$ is the width of the zeroth peak of the autocorrelation of the chaotic time series \cite{le1987conjecture}. In other words, in practice, only $\tau_D/t_c$ values are needed to specify a point on the attractor \cite{lepri1994high}. Even more, it was suggested that a delay can be thought of as a set of $\tau_D/t_c$ roughly independent time slots, such that the $k^{th}$ time slot in one delay is correlated with only the $k^{th}$ time slot in the following delay, as confirmed by the secondary peaks in the autocorrelation function (e.g. Fig. \ref{fig:STR}c). If each of these independent time slots is considered to be a ``node,'' one can think of the delay system as consisting of a set of $\tau_D/t_c$ independent, discrete time nonlinear systems. Clearly, this reasoning is similar to the reasoning that led to the development of the space-time representation and is particularly useful in the same types of situations, i.e., when $\tau_D\gg t_c$.


Temporal discretization arises naturally in many experimental implementations of delay systems. The electro-optic feedback system with a pulsed laser described in Section 2 is one such example \cite{larger2005flow, grapinet2008experimental}. Further, many experimental delay systems implement the delay line with a digital first-in, first-out memory (FIFO) because of the ability to easily vary the delay\cite{murphy2010complex,larger2015laser,hart2016experimental,martinenghi2012photonic,williams2013synchronization,appeltant2011information,hart2017experiments}. In these implementations, the FIFO discretizes time into steps of size $\Delta t=\tau_D/N$, where $N$ is an integer.
These FIFOs apply a constant feedback for one time step $\Delta t$, then sample the system at the end of the time step. Because of the discretization, the use of the co-moving frame $T=\tau_D+\delta$ discussed in Section 2 is not always necessary, and we can simply use a discretized version of original space-time representation Eq. \ref{eq:preSTR}.


In order to reveal the link between these systems and networks, we explicitly discretize time into time steps of length $\Delta t$, and we call each time slot a network node. If $\Delta t$ is chosen to be slightly less than $t_c$, the nodes (which span an interval $\Delta t$) are no longer roughly independent, but are now coupled through the ``inertia'' due to the finite response time of the system to which delayed feedback is applied. This finite response time can be described by a filter impulse response. In this way we have a network of coupled nodes, where the strength and topology of the coupling are determined by the shape of the filter impulse response. The temporal discretization $\Delta t$ is chosen depending on the application, and can have an important impact on the dynamics and coupling, as we discuss at the end of this section.

In order to show explicitly how the network structure arises in these cases, we consider the discretized space-time representation 
\begin{equation}
\label{eq:discreteSTR}
k=nN+i,
\end{equation}
where $k$ is the original discrete time, $n$ is an integer that counts the number of delays that have passed, $N=\tau_D/\Delta t$ is the number of time steps in a delay, and $i$ is the discrete spatial variable. In our network interpretation, $n$ will be the network time and $i$ will be the node index. We impose this discrete space-time representation (Eq. \ref{eq:discreteSTR}) upon the discrete time delayed Eq. \ref{eq:map1}:

\begin{equation}
\label{eq:discrete_convolution}
x^{(i)}[n]=\beta\sum_{m=-\infty}^{nN+i}h[nN+i-m]F(x[m-N]),
\end{equation}
where $N=\tau_D/\Delta t$ is the number of nodes in the network, $n$ is the network time, and $i$ is the node index. We can then split up this summation as follows:

\begin{align}
\label{eq:discrete_sep_sum}
x^{(i)}[n]&=S^{(i)}[n]+C^{(i)}[n] \\
\label{eq:discrete_self}
S^{(i)}[n]&=\beta\sum_{m=-\infty}^{(n-1)N+i}h[nN+i-m]F(x[m-N]) \\
\label{eq:discrete_coupling}
C^{(i)}[n]&=\beta\sum_{m=(n-1)N+i}^{nN+i}h[nN+i-m]F(x[m-N]).
\end{align}

\begin{figure}
\centering
\begin{subfigure}[]{0.45\textwidth}
\includegraphics[width=\textwidth]{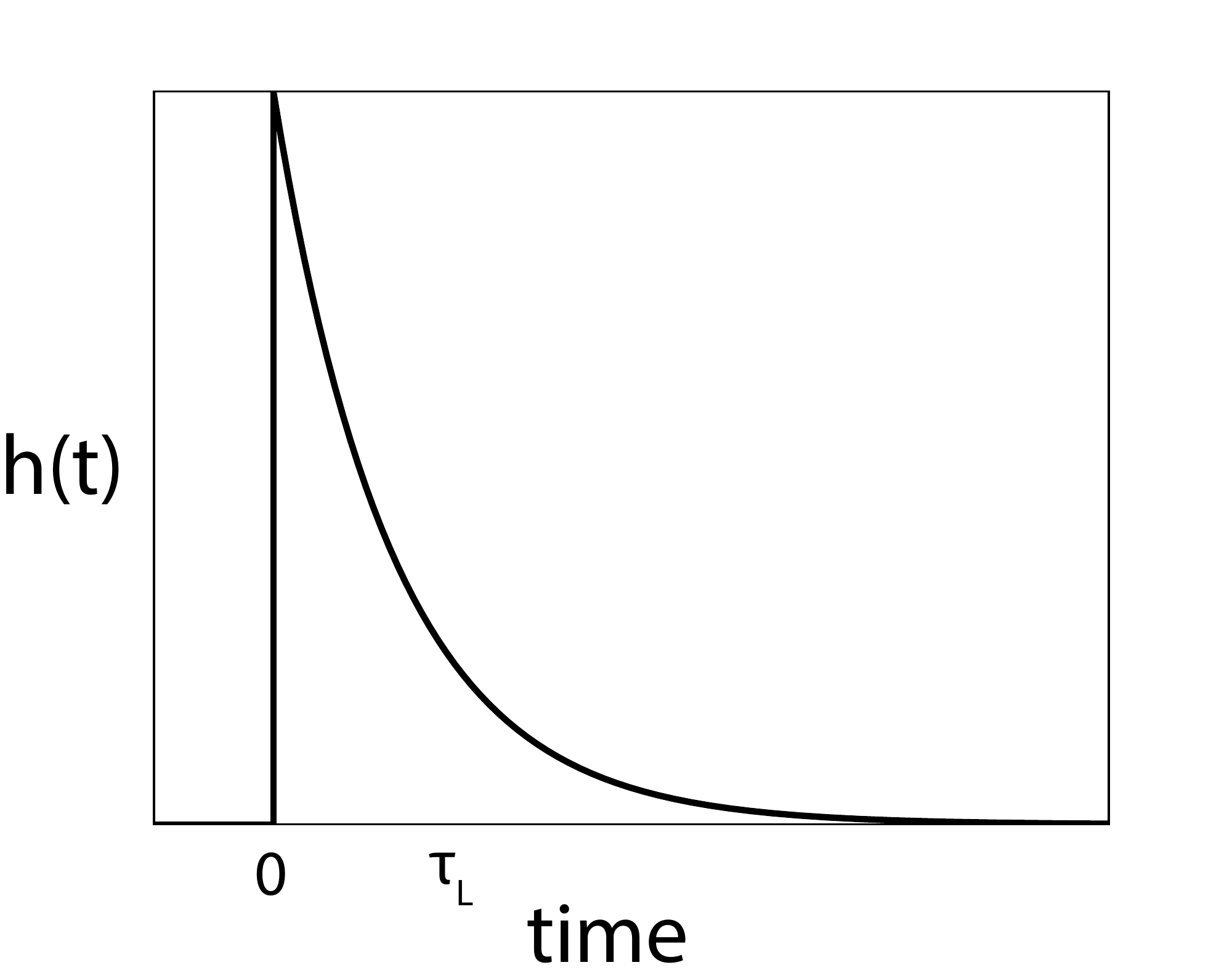}
\caption{\label{fig:LPresponse}}
\end{subfigure}
\begin{subfigure}[]{0.45\textwidth}
\includegraphics[width=\textwidth]{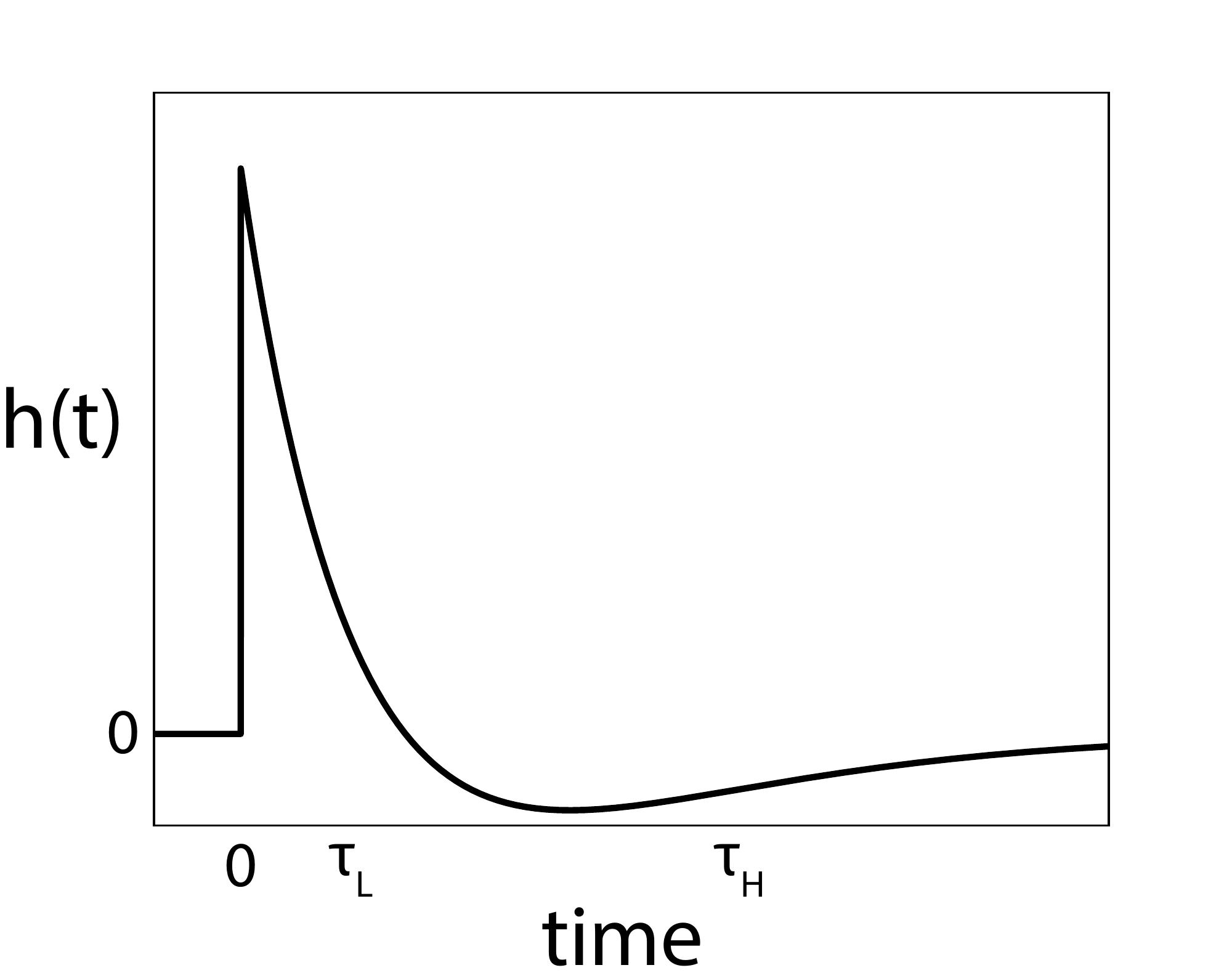}
\caption{\label{fig:BPresponse}}
\end{subfigure}
\caption{Impulse response for (a) single-pole low pass filter and (b) two-pole band pass filter. The poles are real in both cases.}
\end{figure}

Further insight into the meaning of $S^{(i)}[n]$ can be provided by a concrete example. Here we consider the simplest filter, a single-pole low pass filter described by $h(t)=\tau_L^{-1}e^{-t/\tau_L}u(t)$, where $u(t)$ is the Heaviside step function, as depicted in Fig. \ref{fig:LPresponse}. In this case Eq. \ref{eq:discrete_self} becomes

\begin{equation}
\label{eq:discrete_lp_self}
S^{(i)}[n]=\beta e^{-\tau_D/\tau_L} x^{(i)}[n-1].
\end{equation}
Eq. \ref{eq:discrete_lp_self} shows that $S^{(i)}[n]$ is a self-feedback term with a weight $w_h$ that depends on the form of $h(t)$. In general when the delay is long relative to the filter time scales, $w_h\to 0$, as is clear from Eq. \ref{eq:discrete_lp_self} for the particular case of a low pass filter where $w_h=e^{-\tau_D/\tau_L}$.

In order to interpret $C^{(i)}[n]$, we perform a simple change of variables $p=m-nN$ in Eq. \ref{eq:discrete_coupling} to obtain

\begin{equation}
\label{eq:discrete_pre-STR_network}
C^{(i)}[n]=\beta\sum_{p=i+1-N}^{i}h[i-p]F(x^{(p)}[n-1]).
\end{equation}

Therefore $C^{(i)}[n]$ is a coupling term: the summation ``couples'' the values of $x^{(p)}[n-1]$ (weighted by $h$) to the value of $x^{(i)}[n-1]$ to determine $x^{(i)}[n]$.

\begin{figure}[]

\sbox\twosubbox{%
  \resizebox{\dimexpr.9\textwidth-1em}{!}{%
    \includegraphics[height=3cm]{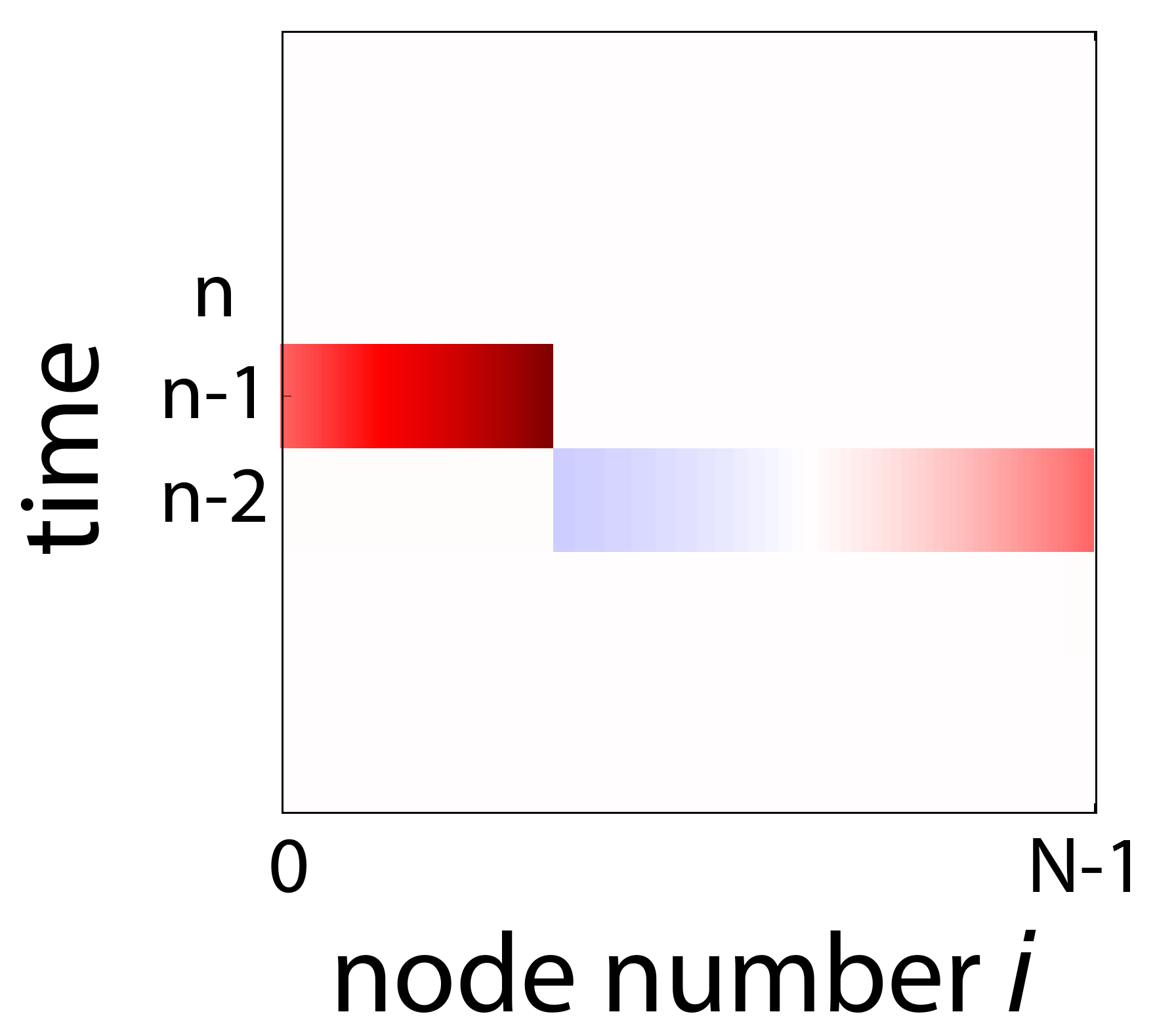}%
    \includegraphics[height=3cm]{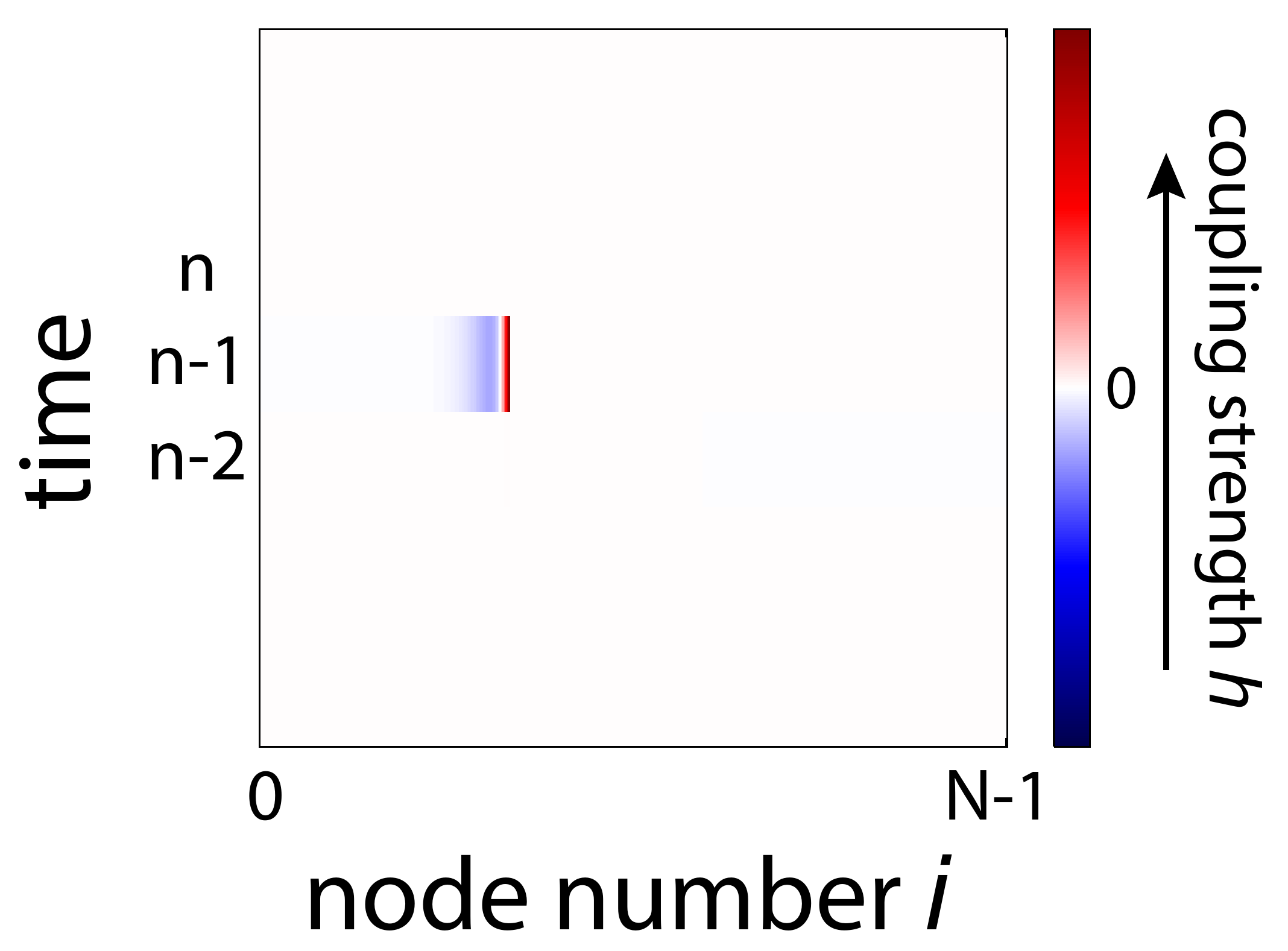}%
  }%
}
\setlength{\twosubht}{\ht\twosubbox}

\centering
\subcaptionbox{}{%
  \includegraphics[height=\twosubht]{STRnetworkbandpass2}%
}\quad
\subcaptionbox{}{%
  \includegraphics[height=\twosubht]{STRnetworkbandpass}%
}

\caption{\label{fig:network_interpretation} Illustration of the coupling term in the space-time representation of delay systems (second term in Eq. \ref{eq:pre-STR_network}) (a) when $\tau_D\approx\tau_L$ and (b) when $\tau_D\gg\tau_L$ when the coupling is implemented by a band pass filter. The coloring indicates the strength of the coupling $h[k]$ from the shaded nodes ($x^{(p)}[n-1]$) to the node represented by the black rectangle ($x^{(i)}[n]$). Red shading represents positive coupling, blue negative coupling, and white no coupling. In (a), the coupling spans two full time steps ($n-1$ and $n-2$), and so this should not be considered a network. In (b), however, the coupling is significant over only a small range (from $p-k_\Delta$ to $p$) and so for almost all nodes $i$ the coupling comes from nodes only at time step $n-1$. Therefore, this can be considered to be a network.}
\end{figure}

Equation \ref{eq:discrete_sep_sum} along with Eqs. \ref{eq:discrete_lp_self} and \ref{eq:discrete_pre-STR_network} now resembles a network equation: each node $i$ is coupled to all the other nodes through the coupling weights $h$. However, this should not yet be considered a network. We recall that the superscript on $x$ denotes a node index and must be in the range [0,N-1]; however, in Eq. \ref{eq:discrete_pre-STR_network} $p$ runs from $i+1-N$ to $i$, which can include negative values. Physically, this means that the coupling summation runs over some $x$ values at time $n-2$ in addition to those from time $n-1$. This is illustrated in Fig. \ref{fig:network_interpretation}a, where the black rectangle denotes $x^{(i)}[n]$ and the shaded region denotes the $x$ values that are coupled to $x^{(i)}[n-1]$ by $C^{(i)}[n]$ to determine $x^{(i)}[n]$.

In cases where the delay $\tau_D=N\Delta t$ is long (relative to the filter time scales), the filter impulse response is significant for only a small range, from $i-k_\Delta$ to $i$, where $k_\Delta\ll N$ is a small number of time steps (determined by the form of $h[k]$) above which $h[k_\Delta]$ is negligible. For long delays, we can approximate Eq. \ref{eq:discrete_sep_sum} as

\begin{equation}
\label{eq:discrete_STRnetwork}
x^{(i)}[n] = w_hx^{(i)}[n-1] +\beta \sum_{p=i-k_\Delta}^{i}h[i-p]F(x^{(p)}[n-1]),
\end{equation}
where the superscript denotes the node number and the number in square brackets denotes the discrete network time. 

Equation \ref{eq:discrete_STRnetwork} is now an exact correspondence with the standard network equation
\begin{equation}
\label{eq:network}
x^{(i)}[n] = G(x^{(i)}[n-1]) +\sum_{j=1}^{N}A_{ij}F(x^{(j)}[n-1]),
\end{equation}
where $G(x)$ is a function that describes the self-feedback and $A_{ij}$ is the weighted network adjacency matrix. By comparing Eqs. \ref{eq:discrete_STRnetwork} and \ref{eq:network}, $G(x)$=$w_hx$. The filter impulse response $h(t)$ is the equivalent of the adjacency matrix; it determines the strength and topology of the coupling.

\begin{figure}[]

\sbox\twosubbox{%
  \resizebox{\dimexpr.9\textwidth-1em}{!}{%
    \includegraphics[height=3cm]{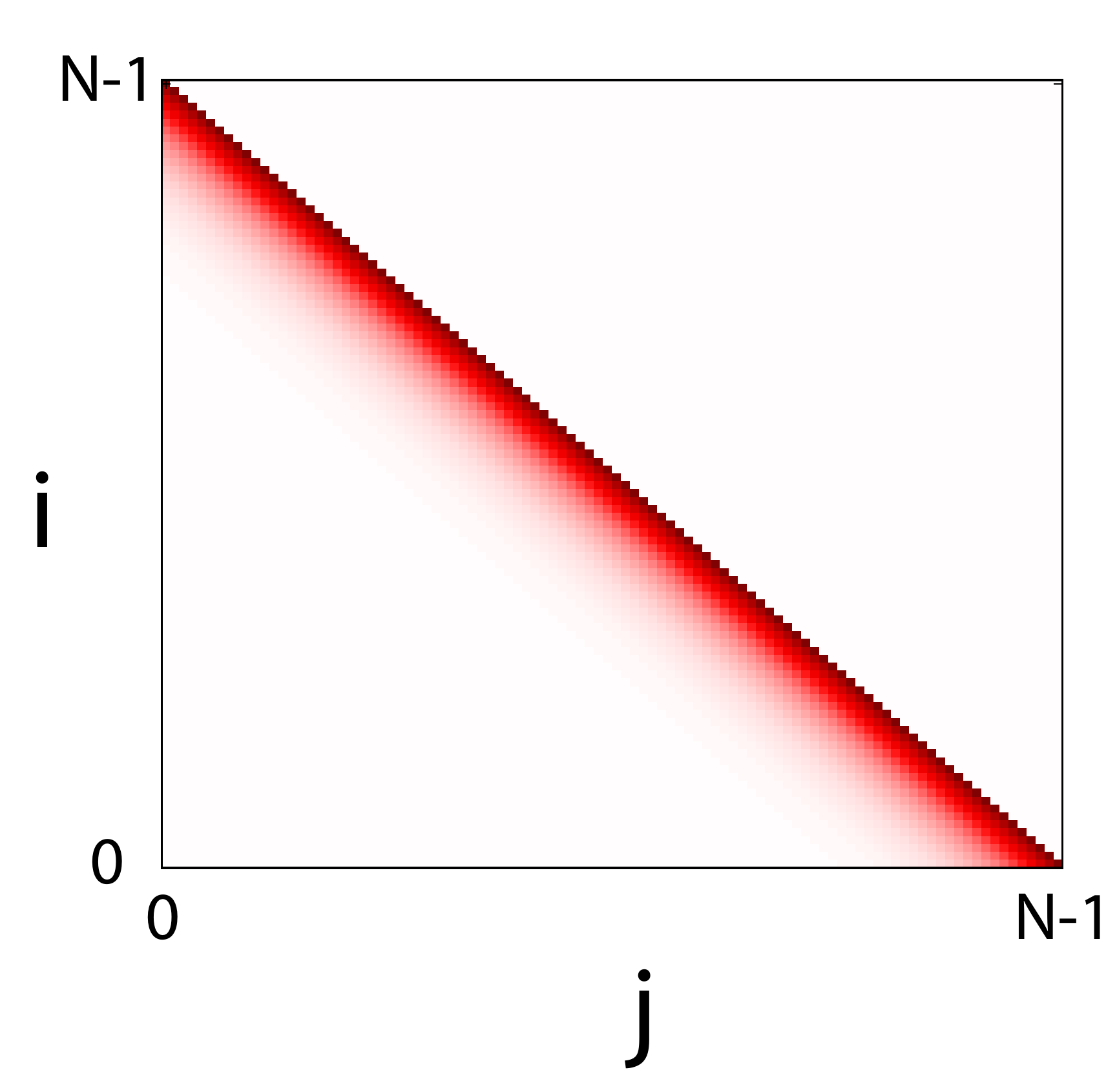}%
    \includegraphics[height=3cm]{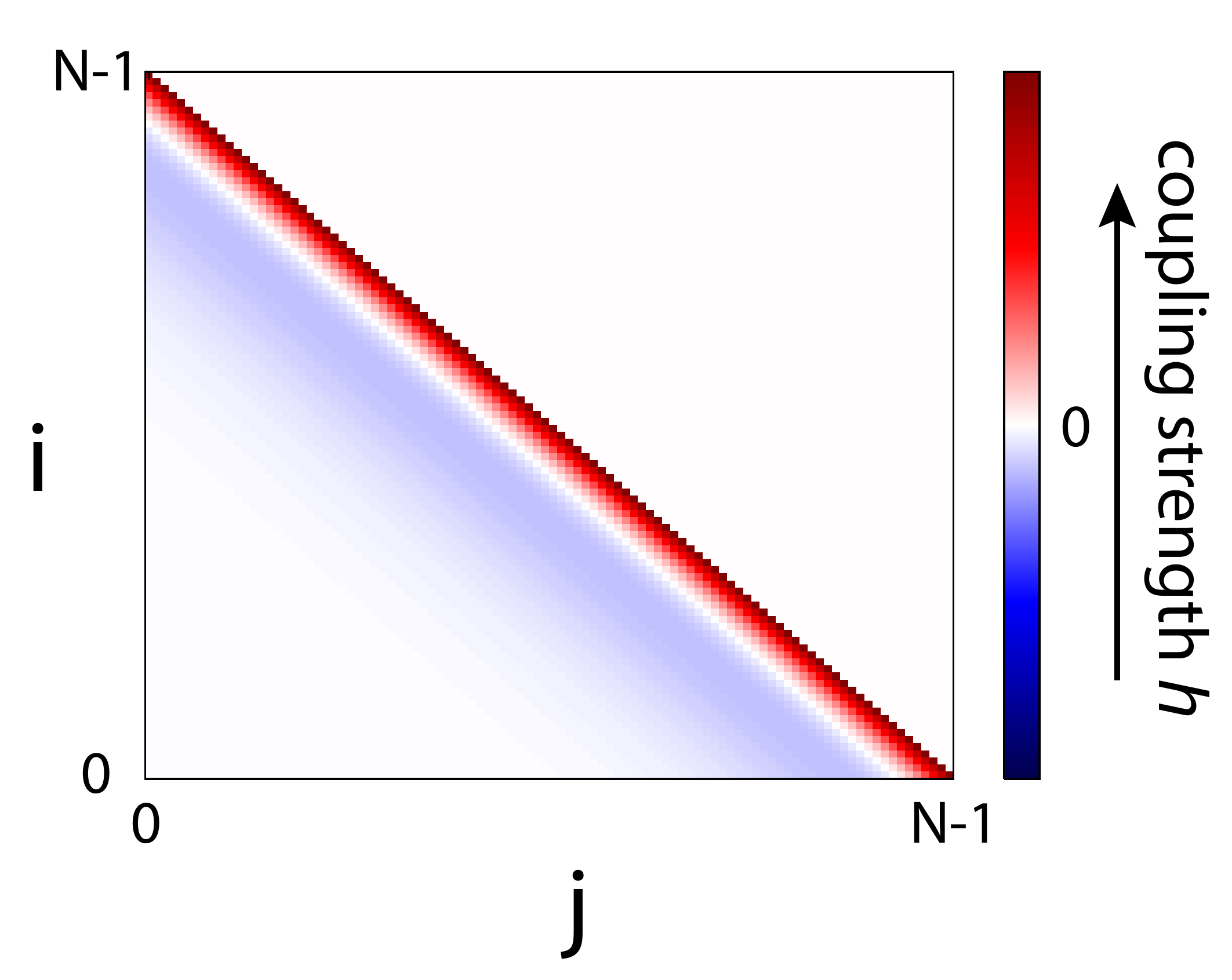}%
  }%
}
\setlength{\twosubht}{\ht\twosubbox}

\centering
\subcaptionbox{\label{fig:lp_adjacency}}{%
  \includegraphics[height=\twosubht]{lp_adjacency}%
}\quad
\subcaptionbox{\label{fig:bp_adjacency}}{%
  \includegraphics[height=\twosubht]{bp_adjacency}%
}

\caption{Illustration of the adjacency matrices for (a) low pass filter and (b) band pass filter. The adjacency matrix is cyclically symmetric due to the time invariance of the filter.}

\end{figure}

For concreteness in demonstration, we now present the adjacency matrices induced by two simple but common impulse responses: the low pass filter and the band pass filter. The single pole low pass filter response is given by \cite{oppenheim1996signals}
\begin{equation}
\label{eq:lp_impulse}
h_{LP}(t) = \tau_L^{-1}e^{-t/\tau_L}u(t),
\end{equation}
where $\tau_L$ is the filter time constant, and $u(t)$ is the Heaviside step function. This is the impulse response that one would use, for example, when solving the Ikeda equation, Eq. \ref{eq:ikeda}. The adjacency matrix that corresponds with this low pass filter is given by

\begin{equation}
\label{eq:lp_adjacency}
  A_{ij}^{LP} = \beta\frac{\Delta t}{\tau_L}
  \begin{cases}
   e^{-(i-j)\Delta t/\tau_L} & \text{if $0\leq i-j\leq k_R$} \\
   0 & \text{otherwise}
  \end{cases}.
\end{equation}
A depiction of this adjacency matrix is shown in Fig. \ref{fig:lp_adjacency}. We note that all couplings are positive and that the network is a directed ring. Another common type of filtering is the two-pole band pass filter, which has impulse response \cite{oppenheim1996signals}

\begin{equation}
\label{eq:bp_impulse}
h_{BP}(t) = \frac{\frac{1}{\tau_L}e^{-t/\tau_L}-\frac{1}{\tau_H}e^{-t/\tau_H}}{1-\tau_L/\tau_H}u(t),
\end{equation}
where $\tau_H$ is the high pass filter time constant and $\tau_L$ is again the low pass filter time constant, depicted in Fig. \ref{fig:BPresponse}. This impulse response corresponds to the filtering in Eq. \ref{eq:DDE}. The corresponding adjacency matrix is 

\begin{equation}
\label{eq:bp_adjacency}
  A_{ij}^{BP} = \beta\frac{\Delta t}{1-\tau_L/\tau_H}
  \begin{cases}
   \tau_L^{-1}e^{-(i-j)\Delta t/\tau_L}-\tau_H^{-1}e^{-(i-j)\Delta t/\tau_H} & \text{if $0\leq i-j\leq k_R$} \\
   0 & \text{otherwise}
  \end{cases}.
\end{equation}
A depiction of this adjacency matrix is shown in Fig. \ref{fig:bp_adjacency}. We note that the network is again a directed ring; however some of the couplings are now negative. Time-invariant filters, such as the two discussed above, will lead to ring networks, and the ring is directed due to causality. However, networks with arbitrary topologies can be created by the introduction of a time dependent filter, as we discuss in Section 7.

Here we make a note about the design of these network experiments and the choice of $\Delta t$ relative to the time scales $\tau_L$ and $\tau_D$. The number of nodes in the network is determined by $\tau_D/\Delta t$; this number should be large for the network interpretation to hold in general. If $\tau_D/\Delta t$ is not large, then the $C^{(i)}[n]$ includes terms from both time $n-1$ and time $n-2$ as shown in Fig. \ref{fig:network_interpretation}a. If $\Delta t<\tau_L$, the (time invariant) filter impulse response will couple the nodes in a cyclically symmetric adjacency matrix, with the coupling radius and coupling strength determined by the form of the impulse response. If $\Delta t \gg \tau_L$, no coupling will be induced by the filtering, and the system will consist of completely independent but identical nodes.

\section{Reservoir computing with delayed feedback}

\begin{figure}
\centering
\includegraphics[width=0.8\textwidth]{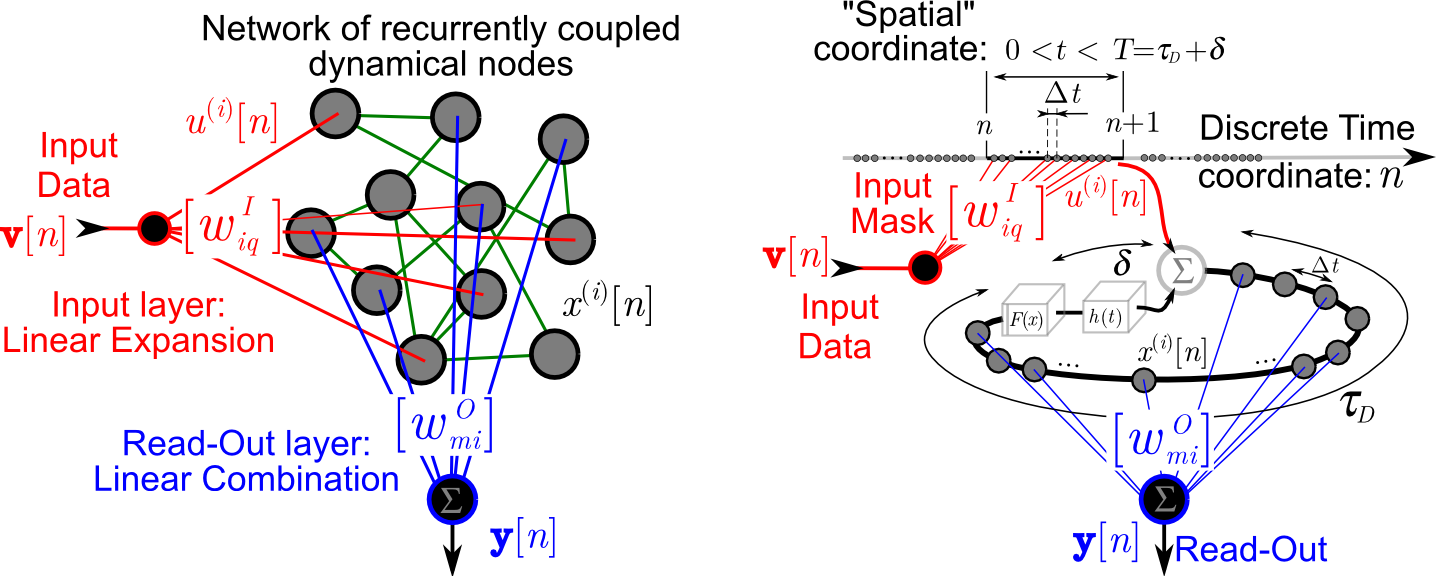}
\caption{\label{fig:RNNvsRC} Graphical comparison between two
  Reservoir Computing implementations: A classical
  RNN architecture (left), and a delay dynamics based
  Reservoir.} 
\end{figure}

Reservoir Computing is a recently proposed brain-inspired processing
technique, corresponding to a simplified version of conventional recurrent neural network (RNN) concepts. It was independently proposed in the machine learning community under the naming Echo State Network (ESN) \cite{jaeger:GMD01} and in the brain cognitive research community as Liquid State Machine \cite{maass:nc02}. It was later unified with the now adopted name, Reservoir Computing (RC) \cite{verstraeten:ipl05, lukosevicius:csr09}. The generic architecture of a RC system is thus rather conventional (see Fig. \ref{fig:RNNvsRC}), consisting of:

\begin{itemize}
\item An input layer aimed at expanding the input information to be
  RC-processed onto each node of the RNN;
\item An internal network having a recurrent connectivity thus
  potentially possessing a complex internal dynamics depending on
  the spectral radius of its connectivity matrix;
\item And an output layer intended to extract the computed result from
  the global observation of the network response, typically performing
  a linear combination of the different internal state variables of
  the network.
\end{itemize}

The most important difference of RC compared to conventional RNN
consists in the restriction of the learning process (i.e. finding the
optimal synaptic weights for the nodes and layer connectivity) to the
output layer only. The input layer and the internal network
connecting weights are usually set at random and are kept
fixed. This makes the learning phase of RC very fast (since reduced to
a linear regression problem), whereas this phase is a difficult and
critical one in RNN, sometimes even not converging. In many
situations, the effective computational power of RC has been found
comparable, or in some cases even better than, their standard RNN counterpart.

One major technological challenge of neuromorphic computing is
however to imagine and design a physical hardware implementing its
specific concepts, instead of translating them into algorithms to be
programmed in standard, however structurally unmatched, digital
processors. The generally recognized poor energy efficiency of
artificial intelligence (AI, involving dedicated supercomputers, or
energy greedy computer farms) is indeed related to the fact that brain
computing concepts have to be adapted into Turing von Neumann
machines, whose architecture and principles of operation are actually
very far from what we have learned from the brain. Up to now
unfortunately, there is essentially no other easily available and
dedicated computing platform capable of efficiently running artificial intelligence techniques.
Turing von Neumann machines are practically the only
effectively working solution today for investigating AI.

An essential problem when one wants to design a dedicated hardware
implementation of neural network processing concepts is the difficulty
to physically fabricate a well controlled three dimensional dynamical
network, as Nature easily does with any brain. Based on
the assumption that what matter are the dynamical complexity and the
high phase space dimension, but not the internal structure itself of the reservoir
network, the EU project PHOCUS (PHOtonic liquid state machine based on
delay CoUpled Systems) started in 2010 with the objective to
demonstrate the RC implementation suitability of nonlinear delay
dynamics. Delay dynamics have thus been proposed as a way to replace a
neural network architecture in the implementation of the RC concepts,
with a first successful demonstration through an electronic delay
system mimicking the Mackey-Glass dynamics \cite{appeltant2011information}. To
do so, extensive use of the space-time analogy of
delay dynamics has been made in order to properly adapt the RC processing rules
previously used in networks of dynamical nodes (and effectively always
programmed or simulated with digital processors).

Figure \ref{fig:RNNvsRC} shows on the left a standard network-based RC
processing (ESN), whereas the right figure displays its analogue based
on nonlinear delayed feedback dynamics for the Reservoir. The
experimental setup first proposed for photonic RC is precisely the one
depicted in Fig.\ref{fig:MZMapparatus}b, in which an external signal is
superimposed at the rf input port of the Mach-Zehnder.\\

\subsection{Input layer}
The input information in standard RNN is expanded into the network
according to spatial multiplexing: The coordinates of the original
input vector $\textbf{v}[n]\in\mathbb{R}^Q$ is expanded through the
multiplication with the input connectivity matrix
$W^I\in\mathbb{R}^N\times\mathbb{R}^Q$. Each node $i=0...N-1$ of the
network is thus receiving an input signal $u^{(i)}[n]$: 

\begin{equation}
  \label{eq:ESN_input}
   u^{(i)}[n] = \sum_{q=1}^Q w^I_{iq}\,v_q[n]
\end{equation}

When one is making use of a delay dynamics instead of network of
nodes, time division multiplexing is naturally adopted to address the
virtual nodes $i$ distributed in time all along the recurrence time
$T$. The required temporal waveform which will need to be injected
into the delay dynamics, reads as follows:

\begin{equation}
  \label{eq:delay_RC_input}
  u(t) = \sum_{i=0}^{N-1} \left[\sum_{q=1}^Q
    w^I_{iq}\,v_q[n]\right]\,p_{\Delta t}(t-nT-i\Delta t),
\end{equation}

\noindent where $p_{\Delta t}(t)$ is a sample and hold function of order
zero. It is a temporal window being unity from time $t=0$ to $t=\Delta t$ and zero everywhere else. 
The duration $\Delta t$ is the sampling
period, or differently speaking, also the temporal spacing between two virtual nodes in the
recurrence time interval $T$. The scalar signal $u(t)$ is practically
programmed in an arbitrary waveform generator, it has the shape of a
piecewise constant signal for each sample $i=0...N-1$ of each time slot of
duration $\Delta t$. When dividing $u(t)$ into sequences of $N$ samples, and
stacking horizontally these vectors of length $N$ for each consecutive
discrete time $n$, one obtains the space-time representation of the
input signal, as depicted in Fig.\ref{fig:STR}c.

\subsection{Reservoir layer}

A transient dynamic is then triggered in the Reservoir due
to the injection of the information signal $u^{(i)}[n]$ or $u(t)$. For the ESN, this
transient is ruled by the following discrete time update rule, from
time $(n-1)$ to time $n$:

\begin{equation}
  \label{eq:ESN_dynamic}
  x^{(i)}[n] = F\left[\sum_{j=1}^N w^R_{ij}\,x^{(j)}[n-1] +
    \rho\cdot u^{(i)}[n]\right],
\end{equation}

\noindent where $W^R\in\mathbb{R}^N\times\mathbb{R}^N$ is the internal
connectivity matrix of the Reservoir. $F[\cdot]$ is a nonlinear
function (usually a sigmoïd, e.g. a hyperbolic tangent, in classical
ESN), and $\rho$ is a scaling factor weighting the input signal
defined in Eq. \ref{eq:ESN_input}.

\noindent In the case of a delay Reservoir, the update rule is similar
to Eq. \ref{eq:convolution}, except the delay dynamics is now
non-autonomous. The input waveform defined in
Eq. \ref{eq:delay_RC_input} is indeed superimposed to the delayed
feedback. It is thus contributing directly to a nonlinear transient in
the delay dynamics phase space, with a contributing weight $\rho$:

\begin{equation}
\label{eq:conv_delayRC}
x(t) = h(t)*F(x(t-\tau_D)+\rho\cdot u(t)) =
\int\limits_{-\infty}^th(t-t')F\big[x(t'-\tau_D)+\rho\cdot u(t)\big]dt'.
\end{equation}

\noindent One could notice that the delay Reservoir, compared to the
discrete time ESN, is continuous in time. The definition of
virtual spatial nodes, and their discretization, is experimentally
introduced through the sampling period $\Delta t$ from Eq. \ref{eq:delay_RC_input}. The adjacency
matrices represented in Figs. \ref{fig:lp_adjacency} and
\ref{fig:bp_adjacency} are then corresponding to the internal
connectivity matrix $W^R$ used for the ESN.\\ The $\Delta t$ time
scale is very important, it has to be properly tuned with respect to the
internal short time $\tau_L$ of the delay dynamics. Optimal processing
efficiency of the delay Reservoir is indeed
empirically found for $\Delta t\simeq\tau_L/5$. This highlights a
necessary comprise between:

\begin{itemize}
\item The definition of sufficiently independent neighboring nodes,
  since $\Delta t$ should not be too short, otherwise adjacent nodes
  are too identical because they are too strongly coupled through the delay
  dynamics inertia (the reservoir response to the input data would
  also be too small in amplitude, since it would be
  strongly filtered; This has detrimental signal-to-noise ratio impacts in
  the RC processing);
\item The other way round, the adjacent nodes could be too decoupled when $\Delta t$
  is too large; If they would be too far one from each other,
  they would allow each stepwise transition of the input
  information to reach an asymptotic state independently of the farther
  past. 
\end{itemize}

\subsection{Output layer}

The last processing operation in RC concerns the Read-Out layer,
consisting of a linear combination of the Reservoir internal states
$x^{(i)}[n]$. This step aims to provide the expected computational
result. The Read-Out operation generates an output vector
$\textbf{y}[n]\in\mathbb{R}^M$, which components read as follows for
the ESN:

\begin{equation}
  \label{eq:Read-Out_ESN}
  y_m[n] = \sum_{i=0}^{N-1} w^O_{mi}\,x^{(i)}[n].
\end{equation}

\noindent The same equation holds in the case of a delay Reservoir,
where however the node state $x^{(i)}[n]$ corresponds to the
extraction of a virtual node state in the delay Reservoir, through the
sampling of $x(t)$. The signal defined by Eq. \ref{eq:conv_delayRC},
is sampled to provide $x(t_k)$, with $t_k=k\cdot\Delta t$, $k$ being
defined as in Eq. \ref{eq:discreteSTR}.

This last processing step of RC involves a learning task, which role
is to determine the coefficients of the linear combination, i.e. the
elements $w^O_{mi}$ of the Read-Out matrix
$W^O\in\mathbb{R}^M\times\mathbb{R}^N$. In the case of supervised
learning, one simply applies a ridge regression to an ill-posed
problem for a set of known data couples, $\{$(\textit{Reservoir
  response}$_l=A_l$, \textit{target Read-Out}$_l=\tilde{B}_l$),
$l=1...L\}$. This corresponds to a training set of $L$ couples of
temporal data (evolution of the discrete time $n$), each having a
duration $N_l$. $A_l\in{\mathbb{R}^N\times\mathbb{R}^{N_l}}$ is thus
the concatenation of the Reservoir state vector
$\{x^{(i)}[n]\,|\,i=0...N-1,\,n=1...N_l\}$, and
$\tilde{B}_l\in\mathbb{R}^M\times\mathbb{R}^{N_l}$ is the same
concatenation for the corresponding target vectors
$\tilde{\textbf{y}}[n]$. The learning requires to consider all
Reservoir responses $A_l$ for the different elements of the training
set, which are gathered into a matrix $A$ (of dimension $N\times(\sum
N_l)$). The latter Reservoir responses are expected to provide the
right corresponding answers $\tilde{B}$ (of dimension $M\times(\sum
N_l)$, $\tilde{B}$ resulting from the concatenation of the target
matrices $\tilde{B}_l$), after having applied to $A$ the searched
Read-Out matrix $W^O$: $\tilde{B}=W^O A$. The ridge regression can be
applied to solve this ill-posed problem, through the following formula
giving the optimal Read-Out matrix:
$W_\text{opt}^O=\tilde{B}\,A^\text{T}\,(A\,A^\text{T}-\lambda
I)^{-1}$, where the superscript $\text{T}$ holds for the matrix
transposition operation, $\lambda$ is the small regression parameter,
$I$ is the $N\times N$ identity matrix, and the matrix inversion can
be calculated through a More-Penrose algorithm.

\begin{figure}
\centering
\includegraphics[width=0.95\textwidth]{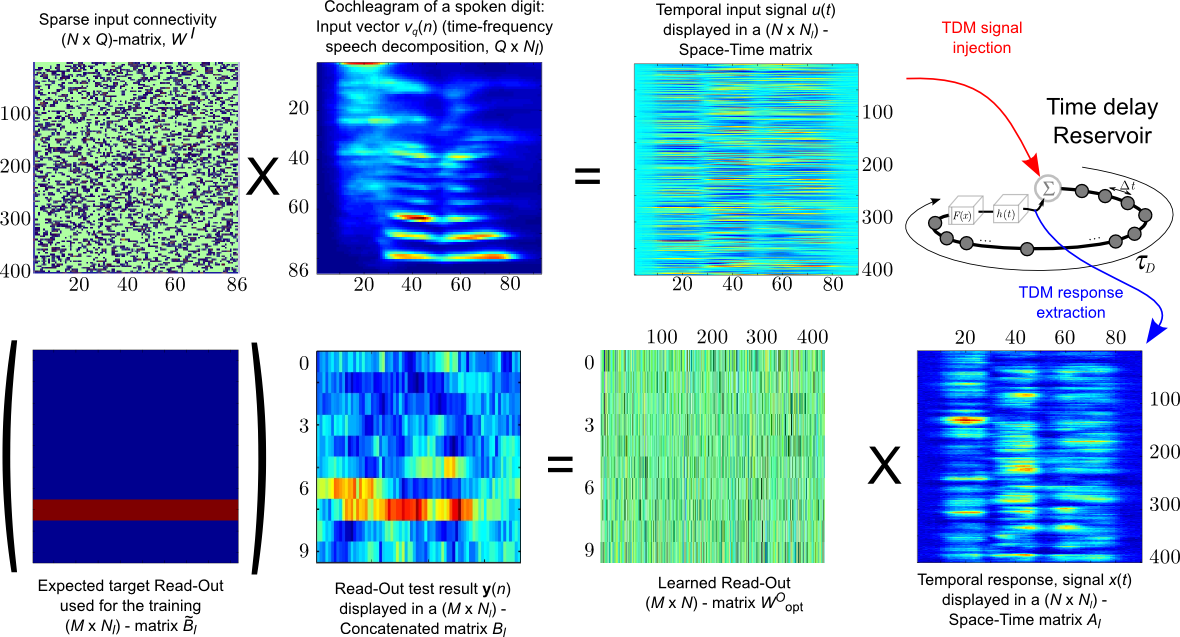}
\caption{\label{fig:RC_proc_SDR} Graphical illustration of the RC
  processing steps in the case of a speech recognition task, performed
  with an optoelectronic delay oscillator used as a Reservoir with 400
  virtual nodes. Each input cochleagram consists of 86 frequency components which energy content (color encoded) are evolving over the duration of the spoken digit (this duration $N_l$ amounts here to 88 steps in $n$).}
\end{figure}

Reservoir Computing has already obtained many successes, revealing its
computational potential both in ESN numerical simulations
\cite{jaeger:sci04,pathak:prl18}, and also in physical hardware
implementation. Successful physical hardware implementations have of course concerned
delay dynamics
\cite{appeltant2011information,larger2012photonic,paquot2012optoelectronic,brunner:ncom13,haynes2015reservoir,larger:prx17},
but also more recently they have been addressed with real spatially
extended photonic systems \cite{brunner:ol15,bueno:ok18}.

Figure \ref{fig:RC_proc_SDR} illustrates the previously described RC
processing steps, in the case the processing of a classification
problem (speech recognition), as performed with an optoelectronic
delay dynamics \cite{larger2012photonic}. It makes an extensive and
illustrative use of the space-time representation for delay dynamical
systems.

\section{The continuum limit}
Networks can also be realized using the space-time representation in the case of fully analog delay lines, such as those that rely on the finite propagation speed of light. Such a system can also be well-approximated by the discrete time systems discussed in Section 3 by taking the limit that $\Delta t/\tau_D\to 0$ \cite{larger2013virtual,larger2015laser}. In these situations, time is continuous, so we return to the space-time representation given by Eq. \ref{eq:STR}. This allows us to think of a continuum of nodes which are labeled by their position $\sigma$ and evolve in discrete time $n$. 


The realization of a network follows very much along the lines of Section 3, but in continuous time rather than discrete. Therefore, the summations will be replaced by integrals, and we will have to account for the drift $\delta$ in the space-time representation. What follows is an elaboration of the presentation contained in Ref. \cite{larger2015laser}.


We begin by analyzing Eq. \ref{eq:convolution} from the perspective of the space-time representation by setting $t = nT + \sigma$ where $n$ is an integer that counts the number of drift-corrected delays $T = \tau_D + \delta$ that have passed since the origin, and $\sigma\in [0,T]$ is the node's position in pseudo-space. 
Re-writing Eq. \ref{eq:convolution} with this change of variables results in 
\begin{equation}
x_n(\sigma)=\beta\int\limits_{-\infty}^{nT+\sigma}h(nT+\sigma- t')F\bigg(x(t'-\tau_D)\bigg)dt'.
\end{equation}
We can then separate the integral into two domains as follows:
\begin{align}
\label{eq:sep_int}
x_n(\sigma)&=S_n(\sigma)+C_n(\sigma)\\
\label{eq:self1}
S_n(\sigma)&=\beta\int\limits_{-\infty}^{(n-1)T+\sigma}h(nT+\sigma- t')F\bigg(x(t'-\tau_D)\bigg)dt'\\
\label{eq:coupling1}
C_n(\sigma)&=\beta\int\limits_{(n-1)T+\sigma}^{nT+\sigma}h(nT+\sigma- t')F\bigg(x(t'-\tau_D)\bigg)dt'.
\end{align}

Further insight into the meaning of $S_n(\sigma)$ can be provided by a concrete example, so that we can evaluate the integral. Here we consider the simplest filter, a single pole low pass filter described by $h(t)=\tau_L^{-1}e^{-t/\tau_L}u(t)$ (Eq. \ref{eq:lp_impulse}). In this case Eq. \ref{eq:self1} becomes
\begin{equation}
\label{eq:lp_self}
S_n(\sigma) = \beta e^{-T/\tau_L}x_{n-1}(\sigma).
\end{equation}
The meaning of $S_n(\sigma)$ is now clear: it is a self-feedback term (from the state $x$ at the spatial position $\sigma$ at discrete time $n-1$ to the state at the spatial position $\sigma$ at discrete time $n$) with a strength determined by the form of $h(t)$.

In order to interpret $C_n(\sigma)$, we make a change of variables $t''=t'+\delta-nT$:
\begin{align}
\label{eq:pre-STR_network}
C_n(\sigma)&=\beta\int\limits_{\sigma - \tau_D}^{\sigma+\delta}h(\sigma + \delta - t'')F\bigg(x_{n-1}(t'')\bigg)dt''.
\end{align}
Therefore $C_n(\sigma)$ is a coupling term: the integral ``couples'' the values of $x_{n-1}(t'')$ to the value of $x_{n-1}(\sigma)$ to determine $x_n(\sigma)$.


When the delay $\tau_D$ is long (relative to the filter time scale), the filter impulse response is significant for only a small range, from $\sigma-\Delta$ to $\sigma+\delta$, where $\Delta\ll\tau_D$ is a short time (determined by the form of $h(t)$) above which $h(t)$ is negligible. For long delays, we can approximate Eq. \ref{eq:pre-STR_network} as 

\begin{equation}
\label{eq:STR_network}
C_n(\sigma)\approx \beta\int\limits_{\sigma - \Delta}^{\sigma+\delta}h(\sigma + \delta - t'')F\bigg(x_{n-1}(t'')\bigg)dt''.
\end{equation}

Eqs. \ref{eq:pre-STR_network} and \ref{eq:STR_network} reveal the network structure that results from viewing the system with long delay through the space-time representation. The system can be interpreted as a continuum of discrete-time nodes whose position (node index) is given by $\sigma$. Each node is coupled to its neighbors within a distance $\Delta$ on the left and $\delta$ on the right through the system's impulse response $h(t)$, as shown in Fig. \ref{fig:network_interpretation}. Importantly, the coupling term in Eq. \ref{eq:STR_network} includes only nodes from time step $n-1$ for almost all nodes $\sigma$ since $\Delta\ll\tau_D$. Indeed, in the limit $\tau_L/\tau_D\to0$, the fraction of nodes whose input coupling spans two time steps vanishes. It is clear from Eq. \ref{eq:STR_network} that $h(t)$ determines both the coupling strength and the coupling width. The particular form of $h(t)$ plays a crucial role in the types of dynamics that the system can exhibit.



\section{Chimeras in systems with delayed feedback}
Chimeras and Reservoir Computing surprisingly share a temporal and a
spatial coincidence. They were ``temporally'' discovered and invented
respectively in the early 2000s
\cite{kuramoto2002coexistence,jaeger:GMD01,maass:nc02}, and they were
``geographically'' connected to delay dynamics during the Delay
Complex System conference DCS'12, a decade later. Since delay dynamics
were successful to demonstrate their capability to emulate a virtual
network of neurons in RC applications, a straightforward challenge was
to also confirm the relevance of this network emulation for the
experimental observation of chimera patterns. Chimeras were moreover
just experimentally found in 2012, in setups modeled by
spatio-temporal equations \cite{hagerstrom2012experimental,tinsley2012chimera}. The
exciting challenge took less than a year until the publication of
delay-based chimera \cite{larger2013complexity}, thus presenting a nice
illustration of delay dynamics efficacy for emulating a network of
dynamical nodes.

Chimera is a particular unexpected solution arising in homogeneous
network of identically coupled oscillators. It manifests itself by a
symmetry breaking solution, since the behavior of the whole
structurally homogeneous network (identical oscillators and coupling
all over the network) splits the network into clusters. Each cluster
coexists one next to the other over long time scales (they consist of a
globally stable solution), and their behavior exhibits coherence
within a cluster, but incoherence between clusters. One of the models
used to numerically explore chimera solutions is the network of
continuously distributed coupled Kuramoto oscillators, defined as
follows: 

\begin{equation}
  \label{eq:Kuramoto_network}
  \frac{\partial{\phi}}{\partial{t}} = \omega_0 + \int G(x-\xi)\cdot
  \sin[\alpha + \phi(t,x) - \phi(t,x-\xi)] \,\text{d}\xi.
\end{equation}

\noindent This governs the dynamics of the phases $\phi(t,x)$ of the
oscillators that are continuously distributed in space, $\omega_0$
being their natural angular frequency. Oscillators have coupled phases
according to a sine nonlinear dependency of the coupling (with an
important coupling offset $\alpha$), depending on the relative phase
difference between the two coupled oscillators at position $x$ and
$x-\xi$. Each phase coupling is weighted by a distance-dependent
factor $G(x-\xi)$, which is typically vanishing beyond a certain
coupling distance (sometimes referred as to the coupling radius)
defined by the shape of $G(\cdot)$. The phase dynamics is thus ruled
by the contribution of the coupling with all the other oscillators, as
the integral term in Eq. \ref{eq:Kuramoto_network} covers the entire
space of the network. Chimera solutions of such an equation typically
consist in clusters, in which oscillators are synchronized with the
same phase in a cluster, and in other clusters, oscillators are
completely
desynchronized with chaotically fluctuating phases.\\
It is then interesting to compare qualitatively the integral term in
Eq. \ref{eq:Kuramoto_network}, with the one derived in
Eq. \ref{eq:STR_network}. As previously discussed and as it can be
also inferred from the comparison with the network of Kuramoto
oscillators, one can clearly identify the specific role of $h(t)$,
when it is considered in the space-time representation of the delay
dynamical variable $x_n(\sigma)$ as derived in Eqs. \ref{eq:sep_int}
to \ref{eq:pre-STR_network}. The impulse response $h(t)$ is clearly ruling the coupling
strength and the coupling distance within the virtual network of
dynamical nodes. The nonlinear function $F(x)$ plays the role of the nonlinear coupling
between the amplitudes of the virtual nodes.

\begin{figure}[h]
\centering
\includegraphics[width=\textwidth]{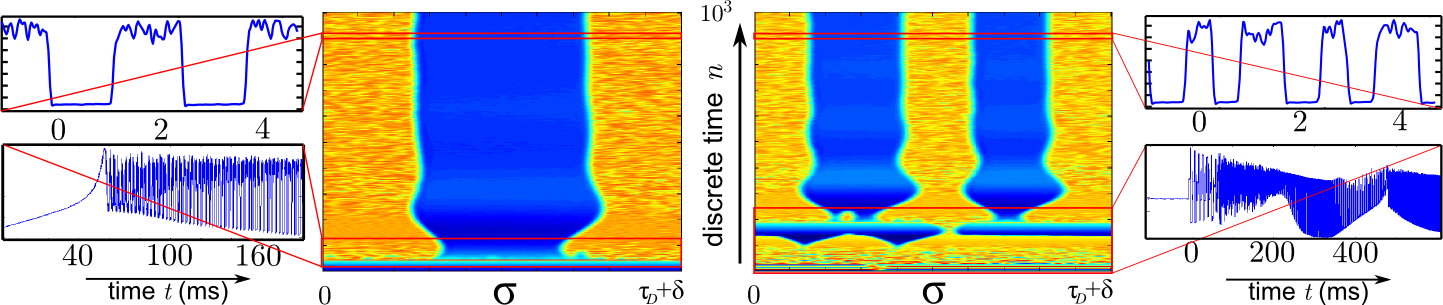}
\caption{\label{fig:chimera_growth} Experimental record of single- and
	two- headed chimera solutions generated in delay dynamics. The two
	central graphs show the space-time representation of the chimera
	solution, as they grow and are then stabilized. The side graphs,
	left and right, are temporal waveforms showing parts of the chimera
	solution, during the initial transient (birth of the chimera from
	the noisy background; lower time-traces, covering a few hundreds of
	time recurrences in the delayed feedback loop), and during the
	stabilized part at the end of the full record (upper time-traces;
	covering approximately two recurrent times $T=\tau_D+\delta$).}
\end{figure}

Figure \ref{fig:chimera_growth} reports typical chimera patterns
obtained experimentally with nonlinear delay dynamics. It shows both
the temporal waveform during growth and stabilization of the pattern,
as well as the space-time representation in the $(\sigma,n)-$plane,
with color encoding of the waveform amplitude. The space-time picture
clearly shows the sustained chimera pattern along the horizontal
virtual space domain. It consists of a flat plateau (blue color)
surrounded by a chaotic sea (red and orange colors), with which it
coexists, filling in a balanced and stable way the shared spatial
domain. The figure also shows actually two possible solutions
(single-headed and two-headed chimera), obtained with the same
parameter conditions, but simply triggered by different noisy initial
conditions. Depending on the temporal parameters (hence the properties
of the coupling function $h(t)$ as depicted in
Fig. \ref{fig:BPresponse}, e.g. the actual values of $\tau_L$ and
$\tau_H$ relatively to $\tau_D$), one can obtain a highly multistable
dynamics of chimera patterns \cite{larger2015laser}. Any $p-$headed
chimera can indeed be generated depending on initial conditions only,
$p$ being any integer below a maximum value fixed by the properties of
$h(t)$.

To comment more into the details under which conditions chimera solutions
can be obtained in delay dynamics, it is worth mentioning that indeed
$h(t)$ requires a bandpass profile. There are many different arguments
to explain this requirement. The first is related to the carrier
waveform of a chimera pattern over the virtual spatial domain
$[n(\tau_D+\delta);(n+1)(\tau_D+\delta)]$, which is necessarily a
stable period-1(delay) carrier waveform, and not a period-2 carrier
waveform as usually concerned in the period-doubling bifurcation
cascade typically known for delay dynamics. To allow for such a
stable period-1 carrier waveform, the bandpass character for $h(t)$ is
necessary (stable period-1 pattern have been analyzed e.g. in
\cite{weicker2013slow}), since the low-pass one is known to lead to
unstable period-1 pattern, as was reported in \cite{giacomelli2012coarsening}
about the ``coarsening'' of any forced initial pattern in the virtual
spatial domain. Last but not least, one could also mention that with a
fixed $\tau_L$, the impulse response with $\tau_H$ (bandpass)
necessarily exhibits a broader width than without the presence of
$\tau_H$ (low-pass). This remark is in line with the known fact that
chimera states are favored when the coupling range is extended
(i.e., beyond the classical case of nearest neighbor coupling only, which
does not allow for chimera states).

\begin{figure}[h]
\centering
\includegraphics[width=0.7\textwidth]{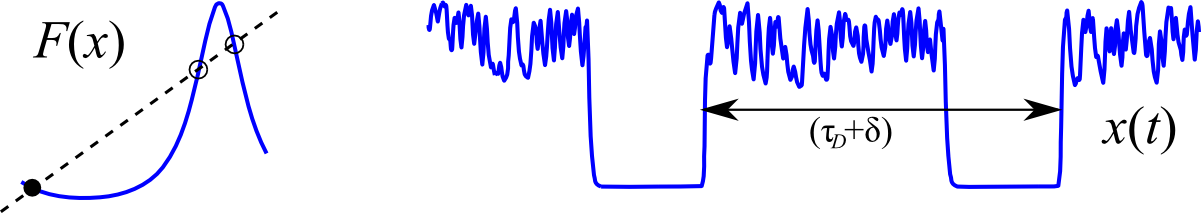}
\caption{\label{fig:chimera_NL} Features of the nonlinear coupling
  function for obtaining chimera patterns in delay dynamics. Left:
  nonlinear function profile $F(x)$, with a dotted first bisector line
  highlighting the fixed points for a map $x_{n+1}=F(x_n)$. Right:
  amplitude correspondance in the temporal chimera waveform $x(t)$.}
\end{figure}

From the point of view of the nonlinear coupling function
between virtual nodes (as the function is involved in
Eq. \ref{eq:STR_network}), there are also specific requirements on
$F(x)$ for obtaining chimera solutions. This
is illustrated in Fig. \ref{fig:chimera_NL}, where both the nonlinear
function profile is represented, and next to it, with the same vertical
scaling, the temporal chimera waveform. From the standard fixed point
analysis for a nonlinear map defined by the same function $F(x)$, one
can notice the following:

\begin{itemize}
\item The nonlinear function operates around an average value centered
  along a positive slope of $F(x)$, between two extrema, where an
  unstable fixed point for the map is located (middle black circle);
\item The high amplitude chaotic part of the chimera waveform
  corresponds to the sharp maximum of $F(x)$, and it develops a
  chaotic motion along this maximum, essentially on the negative slope
  side and centered around an unstable fixed point (upper-right black circle);
\item The low amplitude plateau of the chimera waveform corresponds
  to a stable fixed point (lower-left black disk) of the map, along a
  weak negative slope, thanks to the presence of a broad minimum.
\end{itemize}

This remark points out the important requirement on $F(x)$ about its
necessary asymmetric shape resulting in a sharp maximum and a broad
minimum. This was experimentally obtained in \cite{larger2015laser} with
the Airy function provided by a low finesse P\'erot-Fabry resonator,
which is providing a non linear transformation of the wavelength of a
dynamically tunable laser diode, into the output optical intensity of the
P\'erot-Fabry.

Space-time representation was recently found not to be restricted to a single virtual space dimension. Indeed, adding a second delay much larger than the first one, and acting in parallel to it, enabled 2D chimera to be obtained in delay systems. Among various solutions observed in this two-delay system, one could observe chaotic islands surounded
by a calm sea, or its contrary, a flat plateau island in the middle of a
chaotic sea \cite{brunner:chaos18}.

\section{Arbitrary networks of coupled maps}

Section 3 described the realization of circularly symmetric networks in a single nonlinear system with delayed feedback. In these experiments, the network nodes were time slots of length $\Delta t$, where $\Delta t \ll \tau_D$, and the coupling between nodes was due to the inherent bandwidth of the electronics. This inherent bandwidth was described using a time-invariant infinite impulse response filter; the time invariance results in a circularly symmetric network. However, Eq. \ref{eq:discrete_convolution}  does not require the impulse response to be time-invariant. In this section, we describe recent work that uses a digital filter with a time-varying impulse response to realize arbitrary networks in an experimental delay system \cite{hart2017experiments}.

There are two modifications of previous systems necessary in order to obtain a network with arbitrary topology. \textit{(a)} the inherent circularly symmetric coupling due to the (time-invariant) bandwidth limitations of the system must be removed. \textit{(b)} the desired coupling must be implemented by an appropriately designed filter with a time-dependent impulse.

\begin{figure}
\centering
\begin{subfigure}[]{.45\textwidth}
\includegraphics[width=\textwidth]{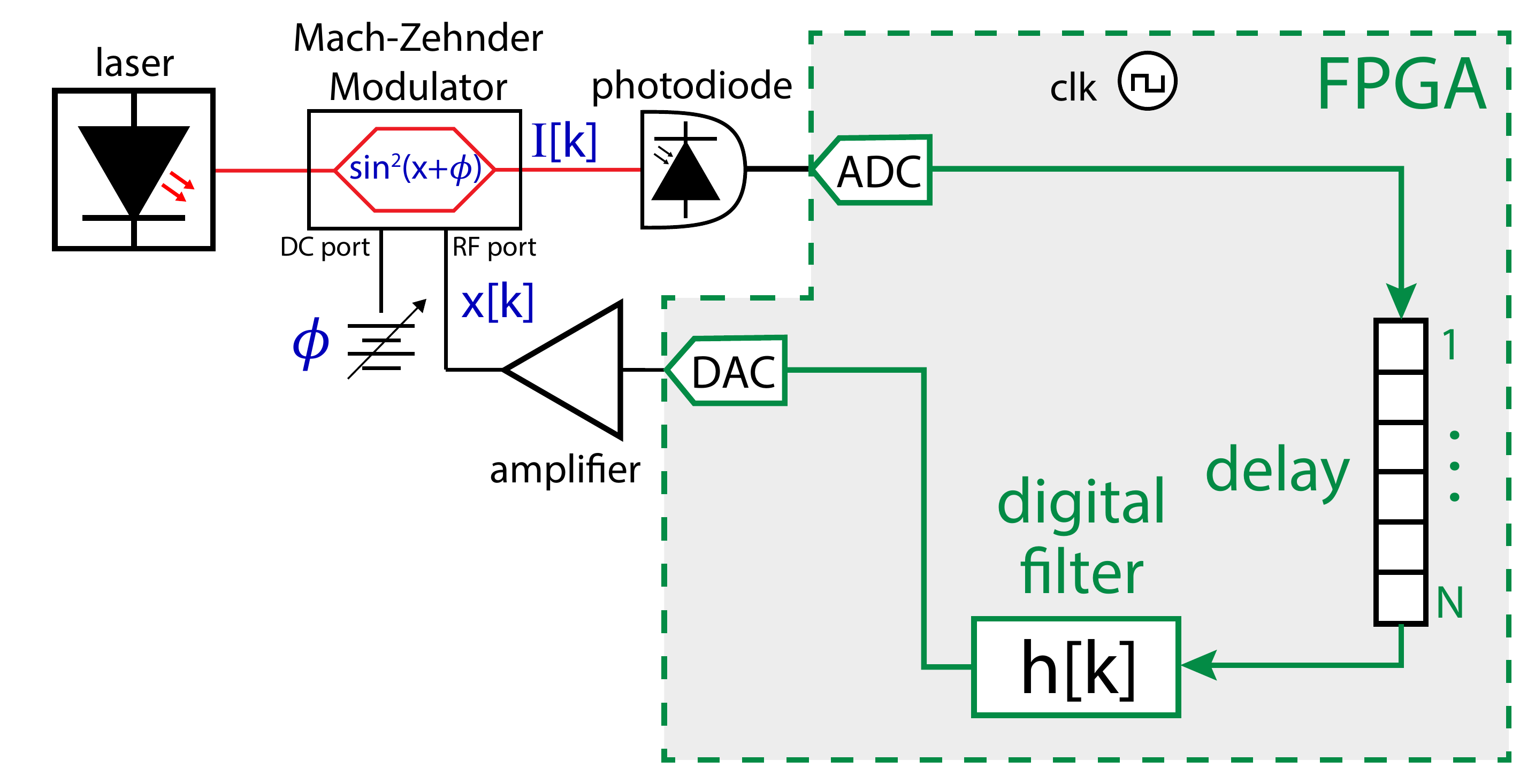}
\caption{}
\end{subfigure}
\begin{subfigure}[]{.45\textwidth}
\includegraphics[width=\textwidth]{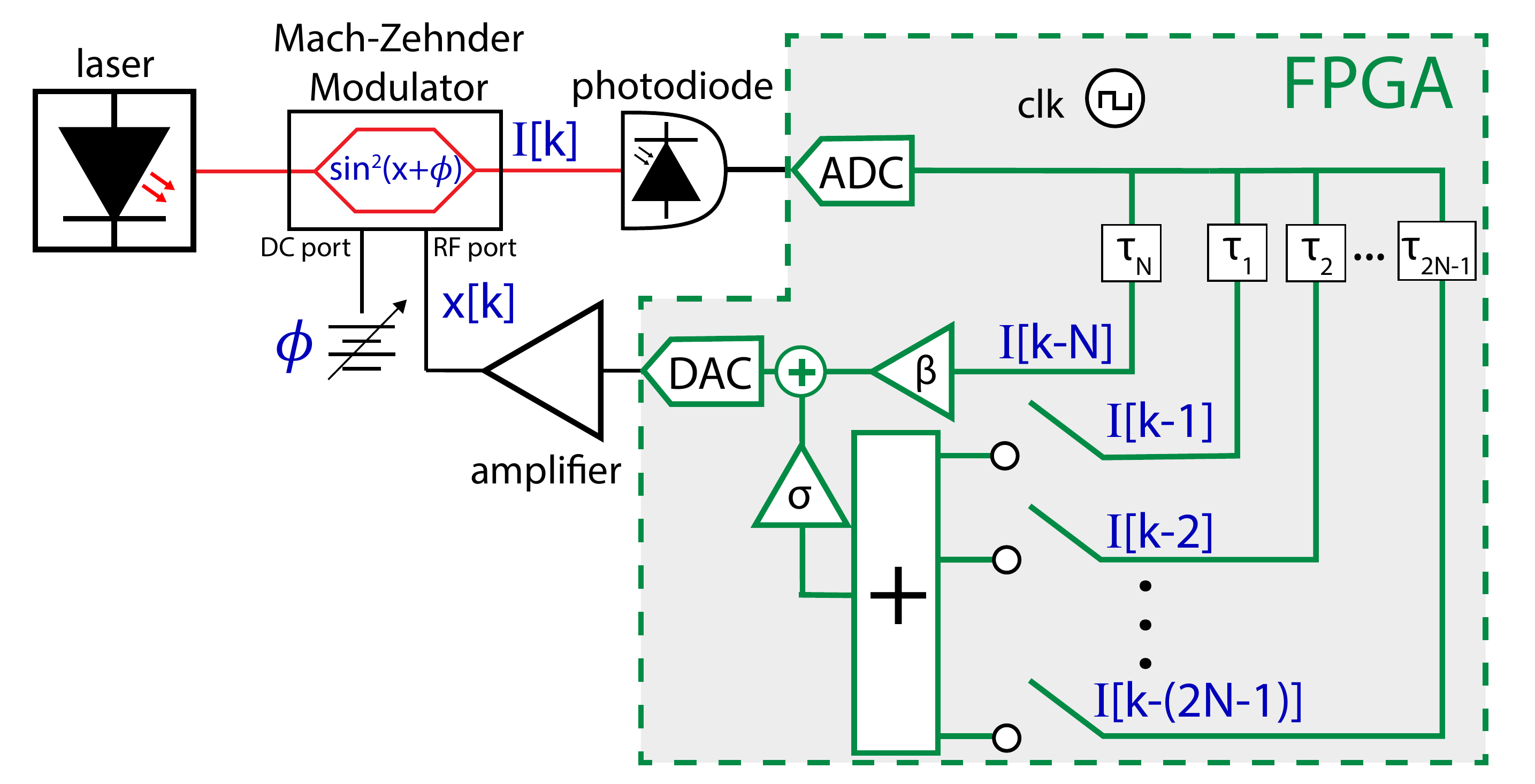}
\caption{}
\end{subfigure}
\caption{\label{fig:ArbMapApparatus}Experimental schematic for realizing arbitrary networks using a single nonlinearity with time multiplexing (a) through a single delay and time-dependent filtering (b) through multiple time-delays that are switched on and off in time. Both illustrations are different ways of viewing the same experiment.}
\end{figure}

\subsection{Removing the inherent circularly symmetric coupling}
There are two convenient options for removing the inherent circularly symmetric coupling due to the time-invariant bandwidth limitations of the system must be removed. 

(I) Perhaps the most straightforward way to remove the coupling due to the bandwidth limitations of the system is to extend the $\Delta t$ described in Section 5. This can be done in the pulsed laser system described by Eq. \ref{eq:map1} by choosing the pulse repetition rate $f_r=N/\tau_D\ll 1/\tau_L$. In this case, the filter response decays before the next pulse arrives, and so the system reduces to the $N$-dimensional map:
\begin{equation}
\label{eq:Nmap}
x[k] = \beta F\big(x[k-N]\big),
\end{equation}
where $k$ is the discrete time.
This map equation requires the specification of $N$ different initial conditions, but the trajectory of each initial condition is completely independent of the trajectories of the others. Therefore, \ref{eq:Nmap} can be thought of as a set of $N$ completely independent but \textit{truly identical} oscillators using the space-time representation:

\begin{equation}
\label{eq:Noscs}
x^{(i)}[n] = \beta F\big(x^{(i)}[n-1]\big),
\end{equation}
where $i=k$ mod $N$ is the oscillator number and $n$ is the network time. 

(II) An easier-to-implement experiment that displays the same map dynamics can be obtained by using a CW laser and sample-and-hold electronics that are clocked at a rate $f_r$. FIFOs, a type of sample-and-hold electronics, have long been used to implement delays in experimental set-ups because of the ease of varying the delay \cite{gibbs1981observation,murphy2010complex,appeltant2011information,martinenghi2012photonic,williams2013synchronization,larger2015laser,hart2016experimental,hart2017experiments}. Such a system can also be described by Eq. \ref{eq:map1}. However, in previous experiments, the clock rates have typically been chosen so that the discrete-time nature of the FIFO delay line minimally impact the dynamics; that is, the sampling time $\Delta t=1/f_r$ has typically been much smaller than any other dynamical time scale, and so the FIFO delay line is a good approximation of an analog delay. In these cases, the experiment is well-described by Eq. \ref{eq:convolution}. Here, we intentionally choose a sampling time that is much longer than the other dynamical time scales in the system, but still shorter than the time delay $\tau_D=N\Delta t$. With this choice of clock rate, the dynamics of the system is well-described by Eq. \ref{eq:Noscs}.

\subsection{Implementing the desired adjacency matrix}

The systems described in the last few paragraphs create $N$ identical, uncoupled nodes using a single delayed dynamical system. In order to couple the nodes together in a network, we must implement a filter that can be described by a time-varying impulse response. This is easiest to do with a digital filter, since in this case we are not restricted by what can be easily implemented by analog components. 

It is convenient to implement both the delay and the digital filter on a single device such as a field-programmable gate array (FPGA). In this case, the filter can be \textit{acausal} in the sense that we can implement the following

\begin{equation}
\label{eq:acausal_impulse}
x[k] = \sum_{m=-\infty}^{(k+N-i-1)}h[k-m;k]F\big(x[m-N]\big),
\end{equation}
where the impulse response $h$ is explicitly written as a function of the discrete time $k$ to denote that it is varying in time. The acausality of the filter is necessary in order to permit couplings to node $i$ from nodes $j>i$.

The impulse response of the digital filter necessary to implement a given network is determined by the adjacency matrix $A_{ij}$ that describes the network as follows:

\begin{equation}
\label{eq:arb_impulse_response}
  h[m;k] = 
  \begin{cases}
   \beta & \text{if $m=k$} \\
   \sigma A_{ij} & \text{if $m\neq k$ and $m=k-i+j$} \\
   0 & \text{otherwise}
  \end{cases},
\end{equation}
where $i=k$ mod $N$ and $j$ is an integer between 0 and $N-1$. 

When the digital filter described by the impulse response in Eq. \ref{eq:arb_impulse_response} is implemented and Eq. \ref{eq:acausal_impulse} is written in the space-time representation, we obtain

\begin{equation}
\label{eq:arbnetwork}
x^{(i)}[n] = \beta F\big(x^{(i)}[n-1]\big) + \sigma \sum_{j}A_{ij}F\big(x^{(j)}[n-1]\big),
\end{equation}

which describes a network of discrete-time oscillators that are coupled by the arbitrary adjacency matrix $A_{ij}$.

There are two adjustments, then, that need to be made to the systems described in Section 4 in order to realize an arbitrary network of coupled oscillators in a single delay system:

\begin{enumerate}[label=(\alph*)]

\item time must be discretized in such a way as to break the nearest-neighbor coupling that would otherwise be induced by the bandwidth limitations of the system. 

\item a filter with a time-dependent impulse response must be used in order to obtain a network topology that is not cyclically symmetric. This filter must also be acausal to allow for the construction of all possible networks (e.g. to couple node $N-1$ to node $0$).
\end{enumerate}

\subsection{Experimental examples}
This technique has been used to implement arbitrary networks in an optoelectronic feedback loop \cite{hart2017experiments}. A schematic of our experiment is shown in Fig. \ref{fig:ArbMapApparatus}a. Light of constant intensity is emitted from a fiber-coupled CW laser. The light passes through an electro-optic intensity modulator, which serves as a nonlinearity. The light is converted to an electrical signal by a photodiode and sampled at a frequency $f_r$ by the FPGA via an analog to digital converter (ADC). The FPGA implements the delay and the time-dependent digital filtering, and outputs the feedback electrical signal through a digital to analog converter (DAC). This signal is amplified and fed back to the modulator, completing the feedback loop.

\begin{figure}
\centering
\includegraphics[width=0.9\textwidth]{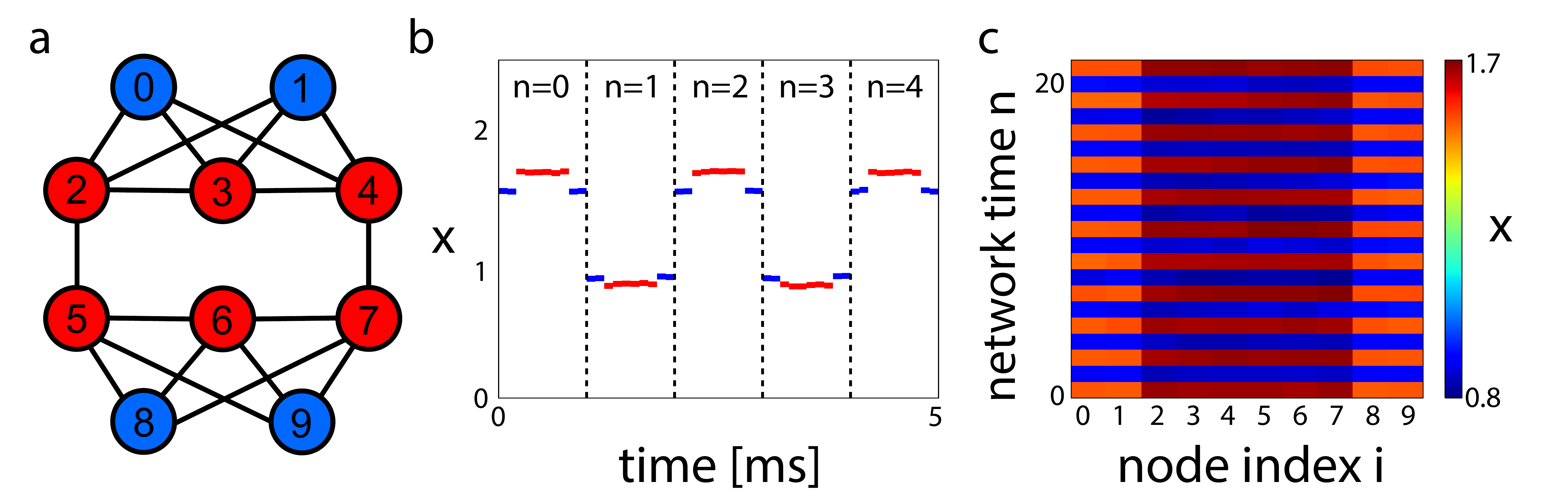}
\caption{\label{fig:ArbNetworkResults}Experimental observation of cluster synchronization using a single-feedback loop implementation of an arbitrary network from ref. \cite{siddique2018symmetry}. (a) Illustration of the network that was implemented. The shading indicates the cluster synchronous state that was observed: nodes that are the same color are in the same synchronous cluster. All nodes are truly identical. (b) Experimentally measured time series. The dotted black lines indicate one network time step. (c) Space-time representation of the time series shown in (b). The cluster synchronous network dynamics are periodic with period two. The parameters used for the measurement are $\beta=1.10$, $\sigma=0.16$, $\phi=\pi/4$.}
\end{figure}

One example of a network that can be implemented using this experimental technique is shown in Fig. \ref{fig:ArbNetworkResults}a. Clearly the network is not rotationally symmetric, so it cannot be implemented by a time-invariant filter. Figure \ref{fig:ArbNetworkResults}b shows experimental time series measured from the system depicted in Fig. \ref{fig:ArbMapApparatus}. If we reorganize this time series according to the space-time interpretation given by Eq. \ref{eq:discreteSTR}, we obtain Fig. \ref{fig:ArbNetworkResults}c, which clearly shows cluster synchronization: nodes 0, 1, 8, and 9 form one synchronized cluster, and nodes 2-7 form the other synchronized cluster.

This network is particularly interesting because it displays an unexpected type of cluster synchronization \cite{siddique2018symmetry}. It had previously been shown that nodes that could be permuted among each other by a symmetry operation could form synchronous clusters \cite{pecora2014cluster}. Later, it was shown that in some cases, symmetry clusters could be combined to form non-symmetric synchronous clusters. This was shown first in Laplacian networks \cite{sorrentino2016complete} then later in more general networks \cite{schaub2016graph}. Figure \ref{fig:ArbNetworkResults}a is a simple example of such a network, as nodes 3 and 6 cannot be permuted with nodes 2, 4, 5, or 7; yet, the red cluster still synchronizes, as shown in Fig. \ref{fig:ArbNetworkResults}c. These experiments confirm the stability of such so-called equitable partition cluster synchronization \cite{siddique2018symmetry}.

\begin{figure}[]
\centering
\includegraphics[width=0.9\textwidth]{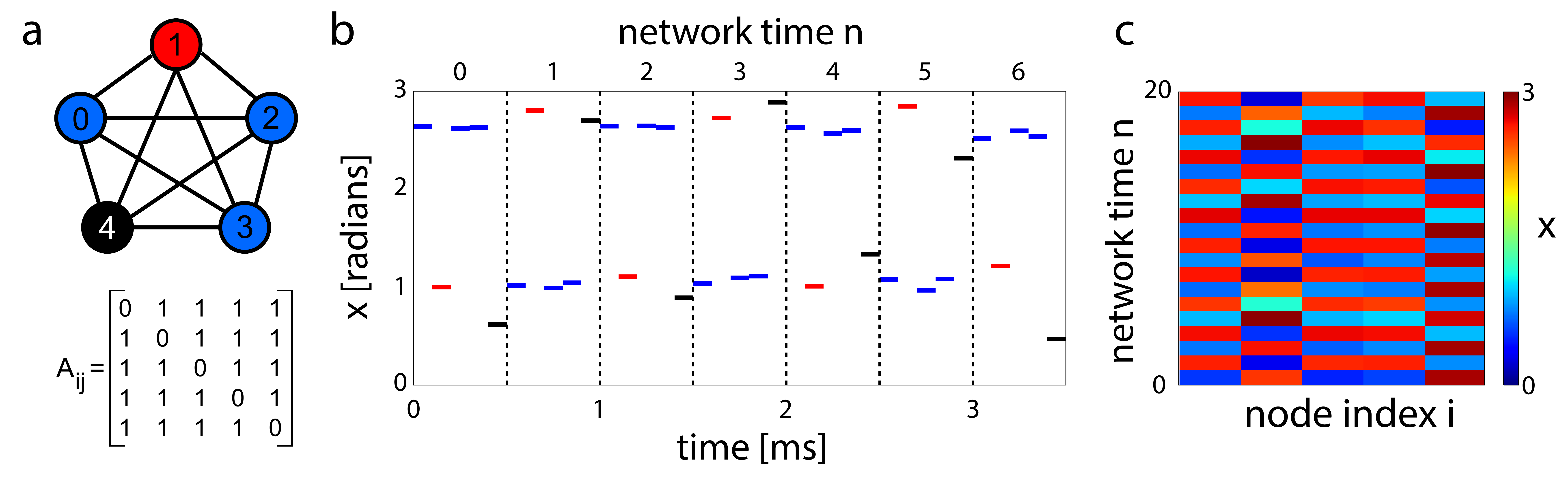}

\caption{\label{fig:5nodechimera}Experimental observation of a chimera state using a single-feedback loop implementation of a globally coupled network. (a) Illustration of the globally-coupled network that was implemented in ref. \cite{hart2017experiments}. The shading indicates the pattern of synchrony that was observed: nodes that are the same color are in the same synchronous cluster. All nodes are truly identical. (b) Experimentally measured time series. The dotted black lines indicate one network time step. (c) Space-time representation of the time series shown in (b). The parameters used for the measurement are $\beta=2.3$, $\sigma=0.23$, $\phi=\pi/4$.}
\end{figure}

As mentioned in Section 7, a chimera state is a dynamical state of a network in which the nodes split up into a coherent set and an incoherent set despite the fact that they are all identical and coupled identically \cite{kuramoto2002coexistence,abrams2004chimera}. The chimeras in Section 7 were observed in a network with circularly symmetric coupling and many nodes. Using the system shown in Fig. \ref{fig:ArbMapApparatus}, we were able to observe a chimera state in a 5 node globally-coupled network \cite{hart2017experiments}. The experimental results are shown in Fig. \ref{fig:5nodechimera}. The globally coupled network and associated adjacency matrix are shown in Fig. \ref{fig:5nodechimera}a. The colors denote the set of synchronized nodes: the blue nodes (0,2,3) are in the coherent set, and the red and black nodes are desynchronized both with the blue nodes and with each other. All nodes are truly identical. Figure \ref{fig:5nodechimera}b shows the time series, where the dotted lines denote the increments of the network time step $n$. Figure \ref{fig:5nodechimera}c shows the space-time representation of the time series, which clearly shows that nodes 0,2, and 3 are synchronized, and nodes 1 and 4 are desynchronized from all nodes. Linear stability calculations confirm that these chimera states are linearly stable \cite{hart2017experiments}.

There is an alternative (but equivalent) way to view the technique used to create arbitrary networks that does not involve acausal filtering. This perspective is described in detail in Ref. \cite{hart2017experiments}. Here, the acausal filter is replaced by multiple delays that are switched on and off as a function of time in order to implement the desired network.  The idea of using multiple time delays to create a more interesting network was pioneered for the purpose of reservoir computing \cite{martinenghi2012photonic}; however, in this case each delay was always switched on, resulting again in a circularly symmetric network (albeit with longer range connections than with a single delay). Switching the additional delays on and off in time breaks the time-invariance (and therefore circular symmetry of the network) and allows an arbitrary network topology. The time-dependent switching is determined according to the following recipe:

\begin{enumerate}
	\item The time delay of length $N$ is always switched ON. This is the feedback time delay and is multiplied by $\beta$. This delay is modeled by the first term in Eq. \ref{eq:arbnetwork}.
	\item Time delays of length $N+i-j$ are switched ON if $A_{ij}=1$, where $i= k$ mod $N$ is the active node. These time delays determine the coupling and are summed then multiplied by $\sigma$. This is modeled by the second term in Eq. \ref{eq:arbnetwork}.
	\item All other time delays are switched OFF.
\end{enumerate}
FIFO time delays and switches are easily implemented in FPGA, making this a particularly powerful implementation because the networks are easy to reconfigure. A schematic of such an experiment is shown in Fig. \ref{fig:ArbMapApparatus}b.


\section{Conclusions and outlook}

The realization of networks of coupled oscillators is a challenging experimental task because of the difficulty and expense of obtaining, coupling, and measuring a large number of identical oscillators. In this paper, we have reviewed recently developed techniques that overcome these obstacles by implementing the network in a single nonlinear delay system through temporal multiplexing. These techniques offer the additional benefit, impossible in other network implementations, that the oscillators are \textit{truly identical} since they are all implemented in the same physical hardware. These delay networks were first developed for their vast potential as a physical implementation of reservoir computing with low cost and high speed. In addition to these important information processing applications, delay networks are also opening up entirely new avenues of research in basic experimental science, as exemplified by the observation of novel 1 and 2 dimensional chimera states and cluster synchronization. These techniques, first conceived only in 2011, are still in their infancy and continue to stimulate basic and applied research.

Future work might explore the use of experimental arbitrary networks for hardware-based reservoir computing, where a time-dependent filter impulse response might allow for the use of a shorter time delay and therefore for faster information processing. This technique can also be used for the experimental study of a variety of fundamental questions of network dynamics, including the impact of targeted perturbations on network dynamics \cite{menck2013basin,menck2014dead}, the effect of heterogeneities on network dynamics \cite{sorrentino2016approximate,nishikawa2016symmetric}, the control of network dynamics \cite{sorrentino2007controllability}, and the impact of noise on network dynamics.

While the delay systems themselves are often continuous time systems, the space-time representation causes delay networks to be discrete in time. Research is currently under way to allow the realization of continuous-time networks in a single delay by adopting the multiple time delay implementation of arbitrary networks, shown in Fig. \ref{fig:ArbMapApparatus}b. Importantly, this technique is not reliant on opto-electronics: one could replace the optics with any system of interest. This might be useful for building prototypes for large networks of coupled oscillators when the oscillators are expensive, such as in the case of power grids. It may also allow for the experimental study of large networks of truly identical oscillators in situations where the oscillators are rarely identical in practice (e.g. biological systems such as neurons). This permits the study of the impact of heterogeneity on the network dynamics.

\textbf{Acknowledgements} 
JDH and RR are supported by the U.S. Office of Naval Research
LL thanks the support from the ANR project BiPhoProc (ANR-14-OHRI-0002-02), and the EIPHI program (ANR-17-EURE-0002).

\bibliographystyle{unsrt}
\bibliography{sample}

\end{document}